\documentclass[%
reprint,
superscriptaddress,
nofootinbib,
 amsmath,amssymb,
 aps 
]{revtex4-2}

\usepackage{multirow}
\usepackage[normalem]{ulem}
\usepackage{booktabs}
\usepackage{graphicx}
\usepackage{dcolumn}
\usepackage{bm}
\usepackage{xcolor}
\usepackage{amsmath}
\usepackage{gensymb}
\usepackage{hyperref}
\usepackage[caption=false]{subfig}
\usepackage{tikz}
\usepackage{orcidlink}
\newcommand{\Albanysuny}{University of Albany, SUNY, Albany, NY 12222, USA}
\newcommand{\Almaty}{Institute of Nuclear Physics at Almaty, Almaty 050032, Kazakhstan
}
\newcommand{\Amsterdam}{University of Amsterdam, NL-1098 XG Amsterdam, The Netherlands}
\newcommand{\Antalya}{Antalya Bilim University, 07190 D\"o{\c s}emealtı/Antalya, Turkey}
\newcommand{\Antananarivo}{University of Antananarivo, Antananarivo 101, Madagascar}
\newcommand{\Antioquia}{University of Antioquia, Medell\'in, Colombia}
\newcommand{\AntonioNarino}{Universidad Antonio Nari\~no, Bogot\'a, Colombia}
\newcommand{\Argonne}{Argonne National Laboratory, Argonne, IL 60439, USA}
\newcommand{\Arizona}{University of Arizona, Tucson, AZ 85721, USA}
\newcommand{\Asuncion}{Universidad Nacional de Asunci\'on, San Lorenzo, Paraguay}
\newcommand{\Athens}{University of Athens, Zografou GR 157 84, Greece}
\newcommand{\Atlantico}{Universidad del Atl\'antico, Barranquilla, Atl\'antico, Colombia}
\newcommand{\Augustana}{Augustana University, Sioux Falls, SD 57197, USA}
\newcommand{\Bern}{University of Bern, CH-3012 Bern, Switzerland}
\newcommand{\Beykent}{Beykent University, Istanbul, Turkey}
\newcommand{\Birmingham}{University of Birmingham, Birmingham B15 2TT, United Kingdom}
\newcommand{\BolognaUniversity}{Universit\`a di Bologna, 40127 Bologna, Italy}
\newcommand{\Boston}{Boston University, Boston, MA 02215, USA}
\newcommand{\Bristol}{University of Bristol, Bristol BS8 1TL, United Kingdom}
\newcommand{\Brookhaven}{Brookhaven National Laboratory, Upton, NY 11973, USA}
\newcommand{\Bucharest}{University of Bucharest, Bucharest, Romania}
\newcommand{\CalBerkeley}{University of California Berkeley, Berkeley, CA 94720, USA}
\newcommand{\CalDavis}{University of California Davis, Davis, CA 95616, USA}
\newcommand{\CalIrvine}{University of California Irvine, Irvine, CA 92697, USA}
\newcommand{\CalLosangeles}{University of California Los Angeles, Los Angeles, CA 90095, USA}
\newcommand{\CalRiverside}{University of California Riverside, Riverside CA 92521, USA}
\newcommand{\CalSantabarbara}{University of California Santa Barbara, Santa Barbara, CA 93106, USA}
\newcommand{\Caltech}{California Institute of Technology, Pasadena, CA 91125, USA}
\newcommand{\Cambridge}{University of Cambridge, Cambridge CB3 0HE, United Kingdom}
\newcommand{\Campinas}{Universidade Estadual de Campinas, Campinas - SP, 13083-970, Brazil}
\newcommand{\CataniaUniversitadi}{Universit\`a di Catania, 2 - 95131 Catania, Italy}
\newcommand{\Catolica}{Universidad Cat\'olica del Norte, Antofagasta, Chile}
\newcommand{\CBPF}{Centro Brasileiro de Pesquisas F\'isicas, Rio de Janeiro, RJ 22290-180, Brazil}
\newcommand{\CEASaclay}{IRFU, CEA, Universit\'e Paris-Saclay, F-91191 Gif-sur-Yvette, France}
\newcommand{\CERN}{CERN, The European Organization for Nuclear Research, 1211 Meyrin, Switzerland}
\newcommand{\Charles}{Institute of Particle and Nuclear Physics of the Faculty of Mathematics and Physics of the Charles University, 180 00 Prague 8, Czech Republic }
\newcommand{\Chicago}{University of Chicago, Chicago, IL 60637, USA}
\newcommand{\ChungAng}{Chung-Ang University, Seoul 06974, South Korea}
\newcommand{\CIEMAT}{CIEMAT, Centro de Investigaciones Energ\'eticas, Medioambientales y Tecnol\'ogicas, E-28040 Madrid, Spain}
\newcommand{\Cincinnati}{University of Cincinnati, Cincinnati, OH 45221, USA}
\newcommand{\Cinvestav}{Centro de Investigaci\'on y de Estudios Avanzados del Instituto Polit\'ecnico Nacional (Cinvestav), Mexico City, Mexico}
\newcommand{\Colima}{Universidad de Colima, Colima, Mexico}
\newcommand{\ColoradoBoulder}{University of Colorado Boulder, Boulder, CO 80309, USA}
\newcommand{\ColoradoState}{Colorado State University, Fort Collins, CO 80523, USA}
\newcommand{\Columbia}{Columbia University, New York, NY 10027, USA}
\newcommand{\conida}{Comisi\'on Nacional de Investigaci\'on y Desarrollo Aeroespacial, Lima, Peru}
\newcommand{\Cti}{Centro de Tecnologia da Informacao Renato Archer, Amarais - Campinas, SP - CEP 13069-901}
\newcommand{\CUSB}{Central University of South Bihar, Gaya, 824236, India
}
\newcommand{\CzechAcademyofSciences}{Institute of Physics, Czech Academy of Sciences, 182 00 Prague 8, Czech Republic}
\newcommand{\CzechTechnical}{Czech Technical University, 115 19 Prague 1, Czech Republic}
\newcommand{\DannecyleVieux}{Laboratoire d'Annecy de Physique des Particules, Universit\'e Savoie Mont Blanc, CNRS, LAPP-IN2P3, 74000 Annecy, France}
\newcommand{\Daresbury}{Daresbury Laboratory, Cheshire WA4 4AD, United Kingdom}
\newcommand{\Dordt}{Dordt University, Sioux Center, IA 51250, USA}
\newcommand{\Drexel}{Drexel University, Philadelphia, PA 19104, USA}
\newcommand{\Duke}{Duke University, Durham, NC 27708, USA}
\newcommand{\Durham}{Durham University, Durham DH1 3LE, United Kingdom}
\newcommand{\Edinburgh}{University of Edinburgh, Edinburgh EH8 9YL, United Kingdom}
\newcommand{\EIA}{Universidad EIA, Envigado, Antioquia, Colombia}
\newcommand{\Eotvos}{E\"otv\"os Lor\'and University, 1053 Budapest, Hungary}
\newcommand{\erciyes}{Erciyes University, Kayseri, Turkey}
\newcommand{\FCULport}{Faculdade de Ci\^encias da Universidade de Lisboa - FCUL, 1749-016 Lisboa, Portugal}
\newcommand{\FederaldeAlfenas}{Universidade Federal de Alfenas, Po{\c c}os de Caldas - MG, 37715-400, Brazil}
\newcommand{\FederaldeGoias}{Universidade Federal de Goias, Goiania, GO 74690-900, Brazil}
\newcommand{\FederaldoABC}{Universidade Federal do ABC, Santo Andr\'e - SP, 09210-580, Brazil}
\newcommand{\FederaldoRio}{Universidade Federal do Rio de Janeiro, Rio de Janeiro - RJ, 21941-901, Brazil}
\newcommand{\Fermi}{Fermi National Accelerator Laboratory, Batavia, IL 60510, USA}
\newcommand{\Ferrarauniv}{University of Ferrara, Ferrara, Italy}
\newcommand{\Florida}{University of Florida, Gainesville, FL 32611-8440, USA}
\newcommand{\Floridastate}{Florida State University, Tallahassee, FL, 32306 USA}
\newcommand{\Fluminense}{Fluminense Federal University, 9 Icara\'i Niter\'oi - RJ, 24220-900, Brazil }
\newcommand{\Genova}{Universit\`a degli Studi di Genova, Genova, Italy}
\newcommand{\Georgian}{Georgian Technical University, Tbilisi, Georgia}
\newcommand{\Granada}{University of Granada \& CAFPE, 18002 Granada, Spain}
\newcommand{\GranSasso}{Gran Sasso Science Institute, L'Aquila, Italy}
\newcommand{\GranSassoLab}{Laboratori Nazionali del Gran Sasso, L'Aquila AQ, Italy}
\newcommand{\Grenoble}{University Grenoble Alpes, CNRS, Grenoble INP, LPSC-IN2P3, 38000 Grenoble, France}
\newcommand{\Guanajuato}{Universidad de Guanajuato, Guanajuato, C.P. 37000, Mexico}
\newcommand{\Harish}{Harish-Chandra Research Institute, Jhunsi, Allahabad 211 019, India}
\newcommand{\Hawaii}{University of Hawaii, Honolulu, HI 96822, USA}
\newcommand{\hkust}{Hong Kong University of Science and Technology, Kowloon, Hong Kong, China}
\newcommand{\Houston}{University of Houston, Houston, TX 77204, USA}
\newcommand{\Hyderabad}{University of  Hyderabad, Gachibowli, Hyderabad - 500 046, India}
\newcommand{\Idaho}{Idaho State University, Pocatello, ID 83209, USA}
\newcommand{\IFIC}{Instituto de F\'isica Corpuscular, CSIC and Universitat de Val\`encia, 46980 Paterna, Valencia, Spain}
\newcommand{\IGFAE}{Instituto Galego de F\'isica de Altas Enerx\'ias, University of Santiago de Compostela, Santiago de Compostela, 15782, Spain}
\newcommand{\ihep}{Institute of High Energy Physics, Chinese Academy of Sciences, Beijing, China}
\newcommand{\Iitk}{Indian Institute of Technology Kanpur, Uttar Pradesh 208016, India}
\newcommand{\Illinoisinstitute}{Illinois Institute of Technology, Chicago, IL 60616, USA}
\newcommand{\Imperial}{Imperial College of Science, Technology and Medicine, London SW7 2BZ, United Kingdom}
\newcommand{\IndGuwahati}{Indian Institute of Technology Guwahati, Guwahati, 781 039, India}
\newcommand{\IndHyderabad}{Indian Institute of Technology Hyderabad, Hyderabad, 502285, India}
\newcommand{\Indiana}{Indiana University, Bloomington, IN 47405, USA}
\newcommand{\INFNBologna}{Istituto Nazionale di Fisica Nucleare Sezione di Bologna, 40127 Bologna BO, Italy}
\newcommand{\INFNCatania}{Istituto Nazionale di Fisica Nucleare Sezione di Catania, I-95123 Catania, Italy}
\newcommand{\INFNFerrara}{Istituto Nazionale di Fisica Nucleare Sezione di Ferrara, I-44122 Ferrara, Italy}
\newcommand{\INFNFrascati}{Istituto Nazionale di Fisica Nucleare Laboratori Nazionali di Frascati, Frascati, Roma, Italy}
\newcommand{\INFNGenova}{Istituto Nazionale di Fisica Nucleare Sezione di Genova, 16146 Genova GE, Italy}
\newcommand{\INFNLecce}{Istituto Nazionale di Fisica Nucleare Sezione di Lecce, 73100 - Lecce, Italy}
\newcommand{\INFNMilanBicocca}{Istituto Nazionale di Fisica Nucleare Sezione di Milano Bicocca, 3 - I-20126 Milano, Italy}
\newcommand{\INFNMilano}{Istituto Nazionale di Fisica Nucleare Sezione di Milano, 20133 Milano, Italy}
\newcommand{\INFNNapoli}{Istituto Nazionale di Fisica Nucleare Sezione di Napoli, I-80126 Napoli, Italy}
\newcommand{\INFNPadova}{Istituto Nazionale di Fisica Nucleare Sezione di Padova, 35131 Padova, Italy}
\newcommand{\INFNPavia}{Istituto Nazionale di Fisica Nucleare Sezione di Pavia,  I-27100 Pavia, Italy}
\newcommand{\INFNPisa}{Istituto Nazionale di Fisica Nucleare Laboratori Nazionali di Pisa, Pisa PI, Italy}
\newcommand{\INFNRoma}{Istituto Nazionale di Fisica Nucleare Sezione di Roma, 00185 Roma RM, Italy}
\newcommand{\INFNRomavergata}{Istituto Nazionale di Fisica Nucleare Roma Tor Vergata , 00133 Roma RM, Italy}
\newcommand{\INFNSud}{Istituto Nazionale di Fisica Nucleare Laboratori Nazionali del Sud, 95123 Catania, Italy}
\newcommand{\Infntorino}{Istituto Nazionale di Fisica Nucleare, Sezione di Torino, Turin, Italy}
\newcommand{\Ingenieria}{Universidad Nacional de Ingenier\'ia, Lima 25, Per\'u}
\newcommand{\Insubria }{University of Insubria, Via Ravasi, 2, 21100 Varese VA, Italy}
\newcommand{\Iowa}{University of Iowa, Iowa City, IA 52242, USA}
\newcommand{\IowaState}{Iowa State University, Ames, Iowa 50011, USA}
\newcommand{\IPLyon}{Institut de Physique des 2 Infinis de Lyon, 69622 Villeurbanne, France}
\newcommand{\IPM}{Institute for Research in Fundamental Sciences, Tehran, Iran}
\newcommand{\IRLPPC}{Particle Physics and Cosmology International Research Laboratory	, Chicago IL,  60637 USA}
\newcommand{\ISTlisboa}{Instituto Superior T\'ecnico - IST, Universidade de Lisboa, 1049-001 Lisboa, Portugal}
\newcommand{\Ita}{Instituto Tecnol\'ogico de Aeron\'autica, Sao Jose dos Campos, Brazil}
\newcommand{\Iwate}{Iwate University, Morioka, Iwate 020-8551, Japan}
\newcommand{\Jacksonstate}{Jackson State University, Jackson, MS 39217, USA}
\newcommand{\Jawaharlal}{Jawaharlal Nehru University, New Delhi 110067, India}
\newcommand{\Jeonbuk}{Jeonbuk National University, Jeonrabuk-do 54896, South Korea}
\newcommand{\Jyvaskyla}{Jyv\"askyl\"a University, FI-40014 Jyv\"askyl\"a, Finland}
\newcommand{\Kansasstate}{Kansas State University, Manhattan, KS 66506, USA}
\newcommand{\Kavli}{Kavli Institute for the Physics and Mathematics of the Universe, Kashiwa, Chiba 277-8583, Japan}
\newcommand{\KEK}{High Energy Accelerator Research Organization (KEK), Ibaraki, 305-0801, Japan}
\newcommand{\KISTI}{Korea Institute of Science and Technology Information, Daejeon, 34141, South Korea}
\newcommand{\Kyiv}{Taras Shevchenko National University of Kyiv, 01601 Kyiv, Ukraine}
\newcommand{\Lancaster}{Lancaster University, Lancaster LA1 4YB, United Kingdom}
\newcommand{\LawrenceBerkeley}{Lawrence Berkeley National Laboratory, Berkeley, CA 94720, USA}
\newcommand{\LIP}{Laborat\'orio de Instrumenta{\c c}\~ao e F\'isica Experimental de Part\'iculas, 1649-003 Lisboa and 3004-516 Coimbra, Portugal}
\newcommand{\Liverpool}{University of Liverpool, L69 7ZE, Liverpool, United Kingdom}
\newcommand{\LosAlmos}{Los Alamos National Laboratory, Los Alamos, NM 87545, USA}
\newcommand{\Louisanastate}{Louisiana State University, Baton Rouge, LA 70803, USA}
\newcommand{\LpBordeaux}{Laboratoire de Physique des Deux Infinis Bordeaux - IN2P3, F-33175 Gradignan, Bordeaux, France, }
\newcommand{\Lucknow}{University of Lucknow, Uttar Pradesh 226007, India}
\newcommand{\Mainz}{Johannes Gutenberg-Universit\"at Mainz, 55122 Mainz, Germany}
\newcommand{\Manchester}{University of Manchester, Manchester M13 9PL, United Kingdom}
\newcommand{\Massinsttech}{Massachusetts Institute of Technology, Cambridge, MA 02139, USA}
\newcommand{\Medellin}{University of Medell\'in, Medell\'in, 050026 Colombia }
\newcommand{\Michigan}{University of Michigan, Ann Arbor, MI 48109, USA}
\newcommand{\Michiganstate}{Michigan State University, East Lansing, MI 48824, USA}
\newcommand{\MilanoBicocca}{Universit\`a di Milano Bicocca , 20126 Milano, Italy}
\newcommand{\MilanoUniv}{Universit\`a degli Studi di Milano, I-20133 Milano, Italy}
\newcommand{\Minnduluth}{University of Minnesota Duluth, Duluth, MN 55812, USA}
\newcommand{\Minntwin}{University of Minnesota Twin Cities, Minneapolis, MN 55455, USA}
\newcommand{\Mississippi}{University of Mississippi, University, MS 38677 USA}
\newcommand{\napoli}{Universit\`a degli Studi di Napoli Federico II , 80138 Napoli NA, Italy}
\newcommand{\Nikhef}{Nikhef National Institute of Subatomic Physics, 1098 XG Amsterdam, Netherlands}
\newcommand{\Niser}{National Institute of Science Education and Research (NISER), Odisha 752050, India}
\newcommand{\Northdakota}{University of North Dakota, Grand Forks, ND 58202-8357, USA}
\newcommand{\Northernillinois}{Northern Illinois University, DeKalb, IL 60115, USA}
\newcommand{\Northwestern}{Northwestern University, Evanston, Il 60208, USA}
\newcommand{\NotreDame}{University of Notre Dame, Notre Dame, IN 46556, USA}
\newcommand{\NoviSad}{University of Novi Sad, 21102 Novi Sad, Serbia}
\newcommand{\Ohiostate}{Ohio State University, Columbus, OH 43210, USA}
\newcommand{\OregonState}{Oregon State University, Corvallis, OR 97331, USA}
\newcommand{\Oxford}{University of Oxford, Oxford, OX1 3RH, United Kingdom}
\newcommand{\PacificNorthwest}{Pacific Northwest National Laboratory, Richland, WA 99352, USA}
\newcommand{\Padova}{Universt\`a degli Studi di Padova, I-35131 Padova, Italy}
\newcommand{\Panjab}{Panjab University, Chandigarh, 160014, India}
\newcommand{\Parissaclay}{Universit\'e Paris-Saclay, CNRS/IN2P3, IJCLab, 91405 Orsay, France}
\newcommand{\Parisuniversite}{Universit\'e Paris Cit\'e, CNRS, Astroparticule et Cosmologie, Paris, France}
\newcommand{\Parma}{University of Parma,  43121 Parma PR, Italy}
\newcommand{\Pavia}{Universit\`a degli Studi di Pavia, 27100 Pavia PV, Italy}
\newcommand{\Penn}{University of Pennsylvania, Philadelphia, PA 19104, USA}
\newcommand{\PennState}{Pennsylvania State University, University Park, PA 16802, USA}
\newcommand{\PhysicalResearchLaboratory}{Physical Research Laboratory, Ahmedabad 380 009, India}
\newcommand{\Pisa}{Universit\`a di Pisa, I-56127 Pisa, Italy}
\newcommand{\Pitt}{University of Pittsburgh, Pittsburgh, PA 15260, USA}
\newcommand{\Pontificia}{Pontificia Universidad Cat\'olica del Per\'u, Lima, Per\'u}
\newcommand{\PuertoRico}{University of Puerto Rico, Mayaguez 00681, Puerto Rico, USA}
\newcommand{\Punjab}{Punjab Agricultural University, Ludhiana 141004, India}
\newcommand{\QMUL}{Queen Mary University of London, London E1 4NS, United Kingdom
}
\newcommand{\Radboud}{Radboud University, NL-6525 AJ Nijmegen, Netherlands}
\newcommand{\Rice}{Rice University, Houston, TX 77005}
\newcommand{\Rochester}{University of Rochester, Rochester, NY 14627, USA}
\newcommand{\Royalholloway}{Royal Holloway College London, London, TW20 0EX, United Kingdom}
\newcommand{\Rutgers}{Rutgers University, Piscataway, NJ, 08854, USA}
\newcommand{\Rutherford}{STFC Rutherford Appleton Laboratory, Didcot OX11 0QX, United Kingdom}
\newcommand{\Salento}{Universit\`a del Salento, 73100 Lecce, Italy}
\newcommand{\santamarta}{Universidad del Magdalena, Santa Marta - Colombia}
\newcommand{\Sapienza}{Sapienza University of Rome, 00185 Roma RM, Italy}
\newcommand{\SergioArboleda}{Universidad Sergio Arboleda, 11022 Bogot\'a, Colombia}
\newcommand{\Sheffield}{University of Sheffield, Sheffield S3 7RH, United Kingdom}
\newcommand{\SLAC}{SLAC National Accelerator Laboratory, Menlo Park, CA 94025, USA}
\newcommand{\Southcarolina}{University of South Carolina, Columbia, SC 29208, USA}
\newcommand{\SouthDakotaSchool}{South Dakota School of Mines and Technology, Rapid City, SD 57701, USA}
\newcommand{\SouthDakotaState}{South Dakota State University, Brookings, SD 57007, USA}
\newcommand{\StonyBrook}{Stony Brook University, SUNY, Stony Brook, NY 11794, USA}
\newcommand{\SURF}{Sanford Underground Research Facility, Lead, SD, 57754, USA}
\newcommand{\Sussex}{University of Sussex, Brighton, BN1 9RH, United Kingdom}
\newcommand{\Syracuse}{Syracuse University, Syracuse, NY 13244, USA}
\newcommand{\Tecnologica }{Universidade Tecnol\'ogica Federal do Paran\'a, Curitiba, Brazil}
\newcommand{\TelAviv}{Tel Aviv University, Tel Aviv-Yafo, Israel}
\newcommand{\TexasAMcollege}{Texas A\&M University, College Station, Texas 77840}
\newcommand{\TexasAMcorpuscristi}{Texas A\&M University - Corpus Christi, Corpus Christi, TX 78412, USA}
\newcommand{\TexasArlington}{University of Texas at Arlington, Arlington, TX 76019, USA}
\newcommand{\Texasaustin}{University of Texas at Austin, Austin, TX 78712, USA}
\newcommand{\Toronto}{University of Toronto, Toronto, Ontario M5S 1A1, Canada}
\newcommand{\Tufts}{Tufts University, Medford, MA 02155, USA}
\newcommand{\Unifesp}{Universidade Federal de S\~ao Paulo, 09913-030, S\~ao Paulo, Brazil}
\newcommand{\UNIST}{Ulsan National Institute of Science and Technology, Ulsan 689-798, South Korea}
\newcommand{\UniversityCollegeLondon}{University College London, London, WC1E 6BT, United Kingdom}
\newcommand{\univkansas}{University of Kansas, Lawrence, KS 66045}
\newcommand{\UNMSM}{Universidad Nacional Mayor de San Marcos, Lima, Peru}
\newcommand{\ValleyCity}{Valley City State University, Valley City, ND 58072, USA}
\newcommand{\Vigo}{University of Vigo, E- 36310 Vigo Spain}
\newcommand{\VirginiaTech}{Virginia Tech, Blacksburg, VA 24060, USA}
\newcommand{\Warsaw}{University of Warsaw, 02-093 Warsaw, Poland}
\newcommand{\Warwick}{University of Warwick, Coventry CV4 7AL, United Kingdom}
\newcommand{\Wellesley}{Wellesley College, Wellesley, MA 02481, USA}
\newcommand{\Wichita}{Wichita State University, Wichita, KS 67260, USA}
\newcommand{\WilliamMary}{William and Mary, Williamsburg, VA 23187, USA}
\newcommand{\Wisconsin}{University of Wisconsin Madison, Madison, WI 53706, USA}
\newcommand{\Yale}{Yale University, New Haven, CT 06520, USA}
\newcommand{\Yerevan}{Yerevan Institute for Theoretical Physics and Modeling, Yerevan 0036, Armenia}
\newcommand{\York}{York University, Toronto M3J 1P3, Canada}

\newcommand{\piplus}{\ensuremath{\pi^+}}
\newcommand{\piminus}{\ensuremath{\pi^-}}
\newcommand{\pizero}{\ensuremath{\pi^0}}

\newcommand{\GeVc}{GeV/$c$~}
\newcommand{\hA}{\textit{hA}}
\newcommand{\hN}{\textit{hN}}

\newcommand{\geant}{\textsc{Geant4}}

\newcommand{\genie}{\textsc{Genie}}

\def\bracketbar{\smash{\hbox{\kern-7pt\raise3pt%
    \hbox{{\tiny(}{\lower1.5pt\hbox{\bf--}}{\tiny)}}}}}

\begin{document}

\preprint{APS/123-QED}
\title{Measurement of Exclusive $\pi^+$--argon Interactions Using ProtoDUNE-SP}

%

\affiliation{\Albanysuny}
\affiliation{\Almaty}
\affiliation{\Amsterdam}
\affiliation{\Antalya}
\affiliation{\Antananarivo}
\affiliation{\Antioquia}
\affiliation{\AntonioNarino}
\affiliation{\Argonne}
\affiliation{\Arizona}
\affiliation{\Asuncion}
\affiliation{\Athens}
\affiliation{\Atlantico}
\affiliation{\Augustana}
\affiliation{\Bern}
\affiliation{\Beykent}
\affiliation{\Birmingham}
\affiliation{\BolognaUniversity}
\affiliation{\Boston}
\affiliation{\Bristol}
\affiliation{\Brookhaven}
\affiliation{\Bucharest}
\affiliation{\CalBerkeley}
\affiliation{\CalDavis}
\affiliation{\CalIrvine}
\affiliation{\CalLosangeles}
\affiliation{\CalRiverside}
\affiliation{\CalSantabarbara}
\affiliation{\Caltech}
\affiliation{\Cambridge}
\affiliation{\Campinas}
\affiliation{\CataniaUniversitadi}
\affiliation{\Catolica}
\affiliation{\CBPF}
\affiliation{\CEASaclay}
\affiliation{\CERN}
\affiliation{\Charles}
\affiliation{\Chicago}
\affiliation{\ChungAng}
\affiliation{\CIEMAT}
\affiliation{\Cincinnati}
\affiliation{\Cinvestav}
\affiliation{\Colima}
\affiliation{\ColoradoBoulder}
\affiliation{\ColoradoState}
\affiliation{\Columbia}
\affiliation{\conida}
\affiliation{\Cti}
\affiliation{\CUSB}
\affiliation{\CzechAcademyofSciences}
\affiliation{\CzechTechnical}
\affiliation{\DannecyleVieux}
\affiliation{\Daresbury}
\affiliation{\Dordt}
\affiliation{\Drexel}
\affiliation{\Duke}
\affiliation{\Durham}
\affiliation{\Edinburgh}
\affiliation{\EIA}
\affiliation{\Eotvos}
\affiliation{\erciyes}
\affiliation{\FCULport}
\affiliation{\FederaldeAlfenas}
\affiliation{\FederaldeGoias}
\affiliation{\FederaldoABC}
\affiliation{\FederaldoRio}
\affiliation{\Fermi}
\affiliation{\Ferrarauniv}
\affiliation{\Florida}
\affiliation{\Floridastate}
\affiliation{\Fluminense}
\affiliation{\Genova}
\affiliation{\Georgian}
\affiliation{\Granada}
\affiliation{\GranSasso}
\affiliation{\GranSassoLab}
\affiliation{\Grenoble}
\affiliation{\Guanajuato}
\affiliation{\Harish}
\affiliation{\Hawaii}
\affiliation{\hkust}
\affiliation{\Houston}
\affiliation{\Hyderabad}
\affiliation{\Idaho}
\affiliation{\IFIC}
\affiliation{\IGFAE}
\affiliation{\ihep}
\affiliation{\Iitk}
\affiliation{\Illinoisinstitute}
\affiliation{\Imperial}
\affiliation{\IndGuwahati}
\affiliation{\IndHyderabad}
\affiliation{\Indiana}
\affiliation{\INFNBologna}
\affiliation{\INFNCatania}
\affiliation{\INFNFerrara}
\affiliation{\INFNFrascati}
\affiliation{\INFNGenova}
\affiliation{\INFNLecce}
\affiliation{\INFNMilanBicocca}
\affiliation{\INFNMilano}
\affiliation{\INFNNapoli}
\affiliation{\INFNPadova}
\affiliation{\INFNPavia}
\affiliation{\INFNPisa}
\affiliation{\INFNRoma}
\affiliation{\INFNRomavergata}
\affiliation{\INFNSud}
\affiliation{\Infntorino}
\affiliation{\Ingenieria}
\affiliation{\Insubria }
\affiliation{\Iowa}
\affiliation{\IowaState}
\affiliation{\IPLyon}
\affiliation{\IPM}
\affiliation{\IRLPPC}
\affiliation{\ISTlisboa}
\affiliation{\Ita}
\affiliation{\Iwate}
\affiliation{\Jacksonstate}
\affiliation{\Jawaharlal}
\affiliation{\Jeonbuk}
\affiliation{\Jyvaskyla}
\affiliation{\Kansasstate}
\affiliation{\Kavli}
\affiliation{\KEK}
\affiliation{\KISTI}
\affiliation{\Kyiv}
\affiliation{\Lancaster}
\affiliation{\LawrenceBerkeley}
\affiliation{\LIP}
\affiliation{\Liverpool}
\affiliation{\LosAlmos}
\affiliation{\Louisanastate}
\affiliation{\LpBordeaux}
\affiliation{\Lucknow}
\affiliation{\Mainz}
\affiliation{\Manchester}
\affiliation{\Massinsttech}
\affiliation{\Medellin}
\affiliation{\Michigan}
\affiliation{\Michiganstate}
\affiliation{\MilanoBicocca}
\affiliation{\MilanoUniv}
\affiliation{\Minnduluth}
\affiliation{\Minntwin}
\affiliation{\Mississippi}
\affiliation{\napoli}
\affiliation{\Nikhef}
\affiliation{\Niser}
\affiliation{\Northdakota}
\affiliation{\Northernillinois}
\affiliation{\Northwestern}
\affiliation{\NotreDame}
\affiliation{\NoviSad}
\affiliation{\Ohiostate}
\affiliation{\OregonState}
\affiliation{\Oxford}
\affiliation{\PacificNorthwest}
\affiliation{\Padova}
\affiliation{\Panjab}
\affiliation{\Parissaclay}
\affiliation{\Parisuniversite}
\affiliation{\Parma}
\affiliation{\Pavia}
\affiliation{\Penn}
\affiliation{\PennState}
\affiliation{\PhysicalResearchLaboratory}
\affiliation{\Pisa}
\affiliation{\Pitt}
\affiliation{\Pontificia}
\affiliation{\PuertoRico}
\affiliation{\Punjab}
\affiliation{\QMUL}
\affiliation{\Radboud}
\affiliation{\Rice}
\affiliation{\Rochester}
\affiliation{\Royalholloway}
\affiliation{\Rutgers}
\affiliation{\Rutherford}
\affiliation{\Salento}
\affiliation{\santamarta}
\affiliation{\Sapienza}
\affiliation{\SergioArboleda}
\affiliation{\Sheffield}
\affiliation{\SLAC}
\affiliation{\Southcarolina}
\affiliation{\SouthDakotaSchool}
\affiliation{\SouthDakotaState}
\affiliation{\StonyBrook}
\affiliation{\SURF}
\affiliation{\Sussex}
\affiliation{\Syracuse}
\affiliation{\Tecnologica }
\affiliation{\TelAviv}
\affiliation{\TexasAMcollege}
\affiliation{\TexasAMcorpuscristi}
\affiliation{\TexasArlington}
\affiliation{\Texasaustin}
\affiliation{\Toronto}
\affiliation{\Tufts}
\affiliation{\Unifesp}
\affiliation{\UNIST}
\affiliation{\UniversityCollegeLondon}
\affiliation{\univkansas}
\affiliation{\UNMSM}
\affiliation{\ValleyCity}
\affiliation{\Vigo}
\affiliation{\VirginiaTech}
\affiliation{\Warsaw}
\affiliation{\Warwick}
\affiliation{\Wellesley}
\affiliation{\Wichita}
\affiliation{\WilliamMary}
\affiliation{\Wisconsin}
\affiliation{\Yale}
\affiliation{\Yerevan}
\affiliation{\York}
\author{S.~Abbaslu} \affiliation{\IPM}
\author{A.~Abed Abud} \affiliation{\CERN}
\author{R.~Acciarri} \affiliation{\Fermi}
\author{L.~P.~Accorsi} \affiliation{\Tecnologica }
\author{M.~A.~Acero} \affiliation{\Atlantico}
\author{M.~R.~Adames} \affiliation{\Tecnologica }
\author{G.~Adamov} \affiliation{\Georgian}
\author{M.~Adamowski} \affiliation{\Fermi}
\author{C.~Adriano} \affiliation{\Campinas}
\author{F.~Akbar} \affiliation{\Rochester}
% ...
\collaboration{The DUNE Collaboration}
\noaffiliation



\date{\today}

\begin{abstract}
We present the measurement of $\pi^{+}$--argon inelastic cross sections using the ProtoDUNE Single-Phase liquid argon time projection chamber in the incident $\pi^+$ kinetic energy range of 500 -- 800 MeV in multiple exclusive channels (absorption, charge exchange, and the remaining inelastic interactions). The results of this analysis are important inputs to simulations of liquid argon neutrino experiments such as the Deep Underground Neutrino Experiment and the Short Baseline Neutrino program at Fermi National Accelerator Laboratory. They will be employed to improve the modeling of final state interactions within neutrino event generators used by these experiments, as well as the modeling of $\pi^{+}$--argon secondary interactions within the liquid argon.  This is the first measurement of $\pi^+$--argon absorption at this kinetic energy range as well as the first ever measurement of $\pi^{+}$--argon charge exchange.

\end{abstract}

\maketitle


\section{Introduction}\label{sec:Intro}
Liquid Argon Time Projection Chamber (LArTPC) neutrino experiments, such as the Short Baseline Neutrino program at Fermi National Accelerator Laboratory and --- in the future --- the Deep Underground Neutrino Experiment (DUNE)~\cite{Abi_2020}, precisely measure energy-dependent neutrino oscillations by observing the charged particles emitted when neutrinos interact with the argon in the detectors. They attempt to identify the flavor of the neutrinos by tagging the charged lepton produced from charged-current interactions and to reconstruct the neutrino energy from all of the observed particles. Key challenges lie in understanding how energy is shared among the emitted particles and how much energy is lost to unobserved neutral or low-energy particles. Final state interactions (FSI) --- of particles within the nucleus immediately following a neutrino interaction --- have a large impact on energy smearing, as energy can be transferred to undetected particles. Additionally, these interactions, along with interactions with the argon of the particles escaping the nucleus following the neutrino interaction (so called ``secondary interactions'' --- SI), can impact the ability to correctly identify the flavor of the incident neutrino. Thus, the modeling uncertainties of FSI and SI limit the resolution and bias the extraction of oscillation parameters but can be constrained by dedicated hadron-nucleus scattering data~\cite{NEUTTune}\cite{T2KOsc}.  For the DUNE oscillation program, understanding $\pi$--argon FSI and SI will be particularly important, as significant portions of $\nu^{\bracketbar}_\mu$ CC events in its detectors will contain a $\pi^\pm$ in the final state --- for example, resonant pion production accounts for roughly 40\% of interactions at DUNE~\cite{DUNEND}.

ProtoDUNE Single-Phase (ProtoDUNE-SP) serves as a testbed to test the technology of DUNE's horizontal-drift Far Detector (FD) module. It also acts as a source for hadron--argon scattering data from its charged particle test beam. In this work, we present the analysis of $\pi^+$--argon interactions with pion kinetic energy between 500 and 800 MeV. Several categories of interactions (depending on their final state signature) were targeted for study within this analysis:
\begin{itemize}
    \item Absorption: no charged pions above 150 MeV/$c$ momentum and no $\pi^0$;
    \item Charge Exchange: no charged pions above 150 MeV/$c$ momentum and at least one $\pi^0$;
    \item Other Interactions\footnote{Note: this will contain contributions from \piplus--to--\piplus~scattering, double charge exchange (\piplus--to--\piminus), and pion production.}: at least one charged pions above 150 MeV/$c$ momentum.
\end{itemize}
While the majority of pions will be produced with lower kinetic energy as a result of the $\nu$-argon interactions seen at DUNE, the energy range covered by this analysis remains an important source of data with which to constrain the FSI models within neutrino event generators. The spread of these models, hinting at the uncertainty of the rates of these interactions --- and thus the importance of this analysis, can be seen within the model comparisons within Reference~\cite{NEUTTune}.

\section{Experimental Apparatus}\label{sec:Experiment}
ProtoDUNE-SP is a LArTPC housed in a cryostat located in the CERN North Area~\cite{PDSPResults}\cite{PDSPDetector}. It collected data from the H4-VLE charged particle beam line  in the Fall of 2018 and cosmic-ray particles until its decommissioning in 2021. The beam instrumentation, described further in Section~\ref{sec:Beamline}, provides event-triggering for the ProtoDUNE-SP detector, as well as particle identification (PID). The analysis within this paper uses roughly 153k $\pi^+$/$\mu^+$ beam triggers from data taken during this time. Note that $\pi^{\pm}$ and $\mu^{\pm}$ are indistinguishable within the beam line PID.     

\subsection{ProtoDUNE-SP Detector}\label{sec:Detector}
\begin{figure}[!ht]
  \centering
    \includegraphics[width=.45\textwidth]{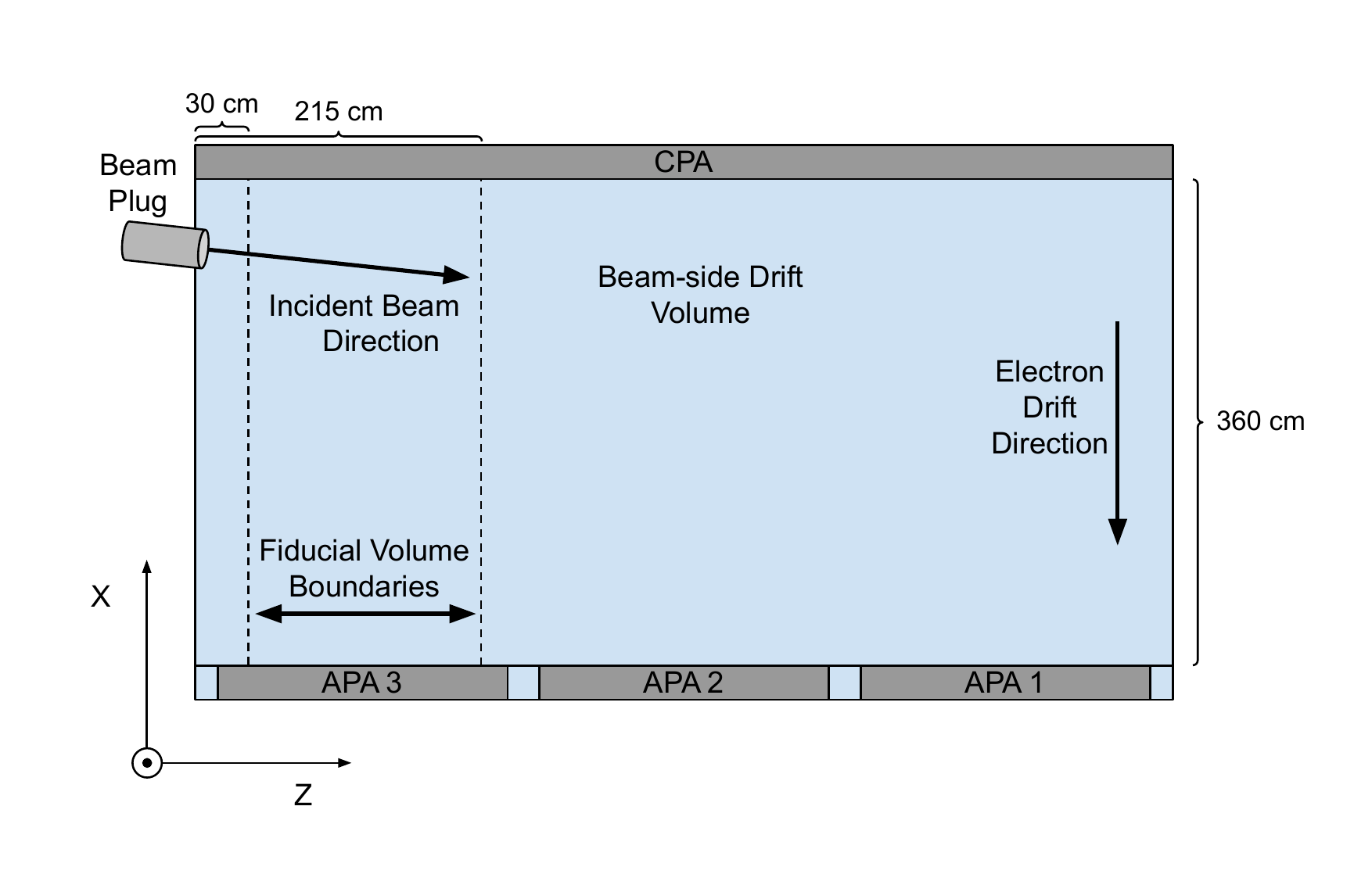}
    \caption{Bird's-eye view of the beam-side drift volume of ProtoDUNE-SP (one half of the full detector). The vertical, dotted line indicates the cut in the reconstructed $z$ position used within the event selection. Along the top of the illustration is the Cathode Plane Assembly (CPA) which separates the two drift volumes. Each beam-side Anode Plane Assembly (APA) is shown at the bottom. Note that the non-beam-side drift volume is not shown as it is not used within the analysis.}
    \label{fig:pdune_sel}
\end{figure}

The ProtoDUNE-SP LArTPC is separated into two drift volumes by a central, vertical cathode plane, with two sets of anode planes on either side of the cathode. Each of the two TPC volumes has dimensions of 3.6 m ($x$) $\times$ 6.1 m ($y$) $\times$ 7.0 m ($z$)\footnote{The directions are defined according to a right-handed coordinate system, where $y$ points upward, and $z$ and $x$ point horizontally. $z$ is parallel to the anode planes.}. The cathode and anode planes are biased such that the nominal drift electric field  magnitude is 500 V/cm. A field cage, which stabilizes the shape of the electric field at the lateral edges of the detector, covers the 4 non-anode faces of the detector. The beam enters into one drift volume by passing through the field cage near the cathode, and points roughly 11$^\circ$ below horizontal and 10$^\circ$ toward the beam-side anode planes. To reduce energy loss upstream of the active TPC volume, a cylindrical ``beam plug'' is installed between the cryostat and the field cage. It is made of alternating fiberglass and stainless steel rings, capped at both ends with fiberglass plates. The plug is filled with nitrogen at 1.3 bar in order to balance against the pressure of the liquid argon while limiting the amount of material through which the test beam particles traverse.  An illustration of the beam-side drift volume of ProtoDUNE-SP is shown in Fig.~\ref{fig:pdune_sel}.

The two sets of anode planes on either side of the cathode are each formed of three Anode Plane Assemblies (APAs) oriented side-by-side. Each APA consists of four planes of wires, three of which are instrumented for readout. The outermost (relative to the center of the APA frame) plane is uninstrumented and provides shielding from long-range induction effects for the sense planes. The wires are biased such that ionization electrons drift past the shielding plane and two instrumented ``induction'' planes before depositing onto the final instrumented ``collection'' plane. The readout planes are oriented $35.7\degree$, $-35.7\degree$, and $0\degree$ with respect to the vertical direction. Thus, combinations of signals on the three wire planes provide 2D positioning in the $yz$ plane of the APA. The third dimension,
$x$,
correlates with the time the charge arrived on the wires. The APAs on the beam-side drift volume were installed with ``electron diverters'' in the gaps between pairs of APAs. These consist of a pair of electrode strips and were intended to prevent ionization from drifting into the region between APAs. However, an electrical short to ground prevented the use of these during operation, and were left unpowered. The unpowered electron diverters absorbed ionization electrons created nearby in the $z$ dimension, resulting in distortion to the tracks in the TPC and loss of charge available for reconstruction. As such, we limit their effect on this analysis (vis-\`a-vis selection performance and energy reconstruction) by separating out tracks which approach the affected region in $z$ from our signal interaction candidates.

The sense wires are read out by a set of cold electronics~\cite{PDSP_CE} which are submerged within the liquid argon at the top of the APAs. The cold electronics shape and amplify the incoming analog signals, which are then digitized at a rate of 2 MHz for processing by the data acquisition system (DAQ). 
When a beam line trigger occurs, the DAQ captures 6000 consecutive samples (3 $ms$) from each wire channel, starting 500 samples (250 $\mu s$) before the trigger, to construct a single event.

\subsection{H4-VLE Beam Line}\label{sec:Beamline}
ProtoDUNE-SP received charged particles from a low energy extension of the H4 beam line called the H4-VLE beam line~\cite{CERN_beam}. 
The beam instrumentation provides ProtoDUNE-SP with event-by-event triggering, momentum reconstruction (using a spectrometer formed of multiple fiber planes which measure the deflection of beam particles by one of the beam magnets),
 and PID using time of flight (TOF) and Cherenkov devices.

ProtoDUNE-SP collected data from $e^+$, $p$, $K^+$, $\pi^+$, and $\mu^+$ in various mean-momentum settings in the range 0.3--7 GeV/$c$. This analysis made use of the $\pi^+$ and $\mu^+$ events in 1 GeV/$c$ momentum runs, which, while the beam line PID was able to separate positron and proton triggers from the $\pi^+$ and $\mu^+$ triggers, were indistinguishable from each other within the beam line PID. The inability to distinguish these particles within the beam line PID had a negligible effect on the analysis, since most of the $\mu^+$ tended to exit the fiducial volume used within the event selection described later on.

\section{Simulation}\label{sec:Simulation}
Events are generated from incident test beam particles resulting from a dedicated G4beamline~\cite{G4beamline} simulation of the 80 GeV/$c$ H4 beam interacting with a 30 cm tungsten target using the FTFP\_BERT \geant~\cite{geant4} physics list. The resulting particles are transported through the H4-VLE extension~\cite{CERN_beam}, and the interactions within the various H4-VLE instrumentation components are simulated. This produces a realistic set of particles which reach the detector and trigger the beam line instrumentation. Cosmic-ray particles are overlaid onto the beam events using the CORSIKA Monte Carlo (MC) Simulation~\cite{Corsika}.

The set of beam and cosmic-ray particles are propagated within the detector volume and their interactions are simulated using \geant with the QGSP\_BERT physics list implemented within LArSoft~\cite{larsoft}. 
The hadronic interaction model within the QGSP\_BERT physics is based on the Bertini intranuclear cascade model~\cite{GUTHRIE196829}\cite{BERTINI1971670} with pre-equilibrium and evaporation models used below an incident hadron energy of roughly 200 MeV~\cite{WRIGHT2015175}.
The ionization of the LAr by the transported charged particles, the subsequent drift of the ionization electrons, and the final detector readout are simulated within LArSoft.
The simulation includes electric field non-uniformities, including those caused by the Space Charge Effect~\cite{uboon_sce} (SCE),
to reproduce more realistically the distortions of the detector response observed in data~\cite{PDSPResults}.
We also include a simulation of the effect of the unpowered electron diverters, in which charge drifting near the gap between APAs is captured.

\subsection{$\pi^\pm$--Nucleus Interaction Modeling}
We compare our results with several cascade models (\geant~Bertini, \geant~INCL++, and \genie~\hN) and one effective model (\genie~\hA). Cascade models share a core principle arising from the fact that, at incident energies relevant to this measurement, the deBroglie wavelengths of hadrons are of the same order or less than the distance between nucleons within a nucleus. As such, individual particle--particle interactions may be simulated within the nucleus~\cite{geant4phys}. Particles are propagated through the nucleus, potentially interacting at each step. These interactions lead to further interactions in the nucleus, and so on, giving rise to a cascade of particles within the nucleus. Once all particles are either absorbed within or escape the nucleus, the cascade ends. Cascade models may differ in specific treatments of the cascades themselves (i.e. the nuclear environment and available constituent interactions) and in additional physics modeling employed to describe aspects of hadron--nucleus interactions outside of the scope of the cascade modeling (i.e. evaporation from excited nuclear states). The models are described as follows:
\begin{enumerate}
    \item \geant~Bertini Cascade Model. This model is based on the intranuclear cascade model originally developed by Bertini~\cite{GUTHRIE196829}\cite{BERTINI1971670}. The implementation within Geant4 is valid from energies of 200 MeV up to $\sim$10 GeV and includes pre-equilibrium physics, nuclear break-up, and evaporation. Below 200 MeV incident energies, the pre-equilibrium model is invoked~\cite{geant4}.
    \item \geant~INCL++ Cascade Model. This is based on a combination of the Li\`ege cascade model~\cite{Liege} and a statistical de-excitation model~\cite{Kelic:2009yg}\cite{Ershova:2023dbv}. Differences to the Bertini cascade model include the description of the nuclear medium, the set of available particle-particle interactions within the nucleus and the associated cross sections, as well as an upper limit on the duration (in time) of the cascade~\cite{geant4phys}. 
    \item \genie~\hN~ Cascade Model. The full intranuclear cascade implemented within the \genie~neutrino event generator samples the angular distributions\footnote{The angular distributions come from partial wave analyses provided by the GWU group~\cite{GWU}.} as functions of energy for several reactions between free $\pi$, $K$, $p$, $n$ and $\gamma$, with adjustments to reflect the influence of the surrounding nuclear medium~\cite{genie}. 
    \item \genie~\hA~ Effective Model. Rather than fully simulating the interactions of individual particles within a nuclear cascade, this effective model uses the total cross section for each possible nuclear process for pions scattering off of iron as a function of energy up to 1.2 GeV. For nuclear targets other than iron, the cross sections are extrapolated by scaling by $A^{2/3}$.
\end{enumerate}

\section{Reconstruction}\label{sec:reco}
Noise filtering, signal processing, and readout issue mitigation are performed on the electronic readout waveforms~\cite{PDSPResults}. Algorithms identify ``hits'' representing individual charge depositions on the readout wires. The Pandora reconstruction software~\cite{Pandora} then performs 2D and then 3D clustering of these hits to identify collections of charge depositions arising from individual particles. Pandora also identifies the primary beam particle and its interaction vertex, and builds the ``tree'' of particles produced from the interaction. A Convolutional Neural Network (CNN)~\cite{PDSPCNN} is used to classify the individual hits as resulting from either track-like particles (i.e. protons, pions, muons) or shower-like particles (i.e. positrons). The reconstructed positions of individual hits are corrected from the distorted readout using data-driven SCE maps~\cite{uboon_sce}. Following calibration and drift electron lifetime corrections, the recorded charge deposited per unit length, $dQ/dX$, of tracks is converted to energy lost per unit length $dE/dX$ using estimates of the electric field strength within the Modified Box model~\cite{ModifiedBox} and these corrected positions~\cite{PDSPResults}\cite{LArPurity}.

\section{Methodology}

In order to measure the cross sections, events in data and from simulation are categorized according to the event selection described in Sec. \ref{sec:Selection}. The simulated event distributions are parameterized and used within a binned likelihood fit to the data as described in Sec. \ref{sec:llhfit}. The best-fit parameters are used to create a varied set of simulated event distributions which best match the data. The cross sections are extracted from truth-level information of the best-fit MC using the thin-slice method developed by LArIAT~\cite{LArIAT_xsec}. We choose to segment the simulated primary \piplus~tracks according to the collection plane wire pitch, though this segmentation is ultimately an arbitrary choice for our application of this method. 

\subsection{Event Selection}\label{sec:Selection}
The analysis made use of beam events which were categorized as $\pi^+/\mu^+$ by the beam instrumentation by requiring no signal in the low-pressure Cherenkov device and a measured TOF less than 110 ns~\cite{PDSPResults} from all 1~\GeVc runs. Events with degenerate hits in any of the scintillating fiber planes of the momentum spectrometer were rejected to avoid ambiguities in momentum reconstruction.

The reconstructed beam momentum spectrum of the data was found to be wider than in MC. The shape difference was accounted for by normalizing MC to data in bins of reconstructed beam momentum. Due to low amounts of events in the tails, the reconstructed beam momentum region was restricted to (750, 1250) MeV/$c$.
As explained in Ref.~\cite{KaonMeasurement}, pre-selection criteria were used to increase the purity and reconstruction quality of selected events. First, there must be a valid reconstructed primary particle: Pandora must have identified a cluster
as coming from the beam, and reconstructed that cluster as track-like as opposed to shower-like. Second, the reconstructed beam track must have extended at least 30 cm into the TPC in the $z$ direction. Finally, a set of positional and directional cuts on the reconstructed beam track were used in order to reduce the number of cosmic rays or interactions upstream of the TPC which were misidentified as the primary beam particle.
The events passing this cut were then placed into one of the following four categories: escaping tracks, absorption candidates, charge exchange candidates, and other interaction candidates.

Escaping tracks are any reconstructed beam track which extended at least 215 cm into the detector in the $z$-direction. The location of this cut in $z$ is chosen such that tracks which are distorted by the grounded electron diverters are not considered interaction candidates. Additionally, including this category in the fit helps to constrain both the relative number of $\mu^+$ in the sample (since $\mu^+$ are more likely to exit the fiducial volume) and the total inelastic cross section. This last point results from the fact that $\pi^+$ will be more (less) likely to travel past this cut in $z$ if their total cross section is lower (higher). Note that this does limit the efficiency of selecting low-kinetic-energy interactions, leading to the restricted low-end of the measurement range (500--800 MeV). 

Reconstructed beam tracks that end before the previous cut in $z$ are all considered interaction candidates and are further separated into the three signal interactions (as described in Section~\ref{sec:Intro}). This is achieved by using the reconstructed particles that are associated with the interaction products of the primary reconstructed beam track. 
A reconstructed interaction product is given a charge-weighted track-like score $T$ as defined in Eq.~\ref{eq:trackscore},
\begin{equation}\label{eq:trackscore}
    T = \frac{1}{Q}\sum\limits_{i}^{N_\textrm{Hits}}t_iq_i;\;
    Q = \sum\limits_i^{N_\textrm{Hits}}q_i,
\end{equation}
where $t_i$ and $q_i$ are the track score from the aforementioned CNN and integrated charge of each hit on the collection plane, and $Q$ is the total charge deposited on the collection plane for the given reconstructed cluster.
We use the charge-weighted score because it yields better agreement between data and simulation in this analysis compared to simply summing individual scores and normalizing over the number of hits. For the same reason, the sum is restricted to hits on the collection plane.

The $dE/dX$ of reconstructed tracks is then used to identify $\pi^\pm$ candidates. Candidates for $\gamma$ resulting from the decay of secondary $\pi^0$s are identified using the energy of the shower and the distance between the end of the beam track and the start of the shower. If a primary beam track has no $\pi^\pm$ candidates and no $\gamma$ candidates in its set of interaction products, it is considered an absorption event. If it has no $\pi^\pm$ candidates and at least one $\gamma$ candidate, it is considered charge exchange. Finally, if it has at least one $\pi^\pm$ candidate, it is considered an ``other" interaction.

\begin{figure}[!ht]
  \centering
 \subfloat[\label{fig:abs_evdisp}]{
    \begin{tikzpicture}
    \node[inner sep=0pt] (image) at (0,0){\includegraphics[width=.45\textwidth]{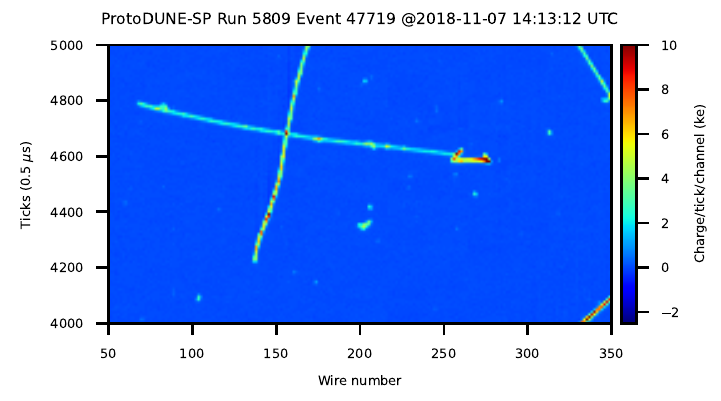}};
    \draw[-latex, thick, red] (-2.5, 1.2) to (1.05,0.7);
    \draw[dashed, thick, red] (1.3, 0.5) ellipse (0.35cm and 0.15cm);
    \end{tikzpicture}
  }
  
  \subfloat[\label{fig:cex_evdisp}]{
    \begin{tikzpicture}
    \node[inner sep=0pt] (image) at (0,0){\includegraphics[width=.45\textwidth]{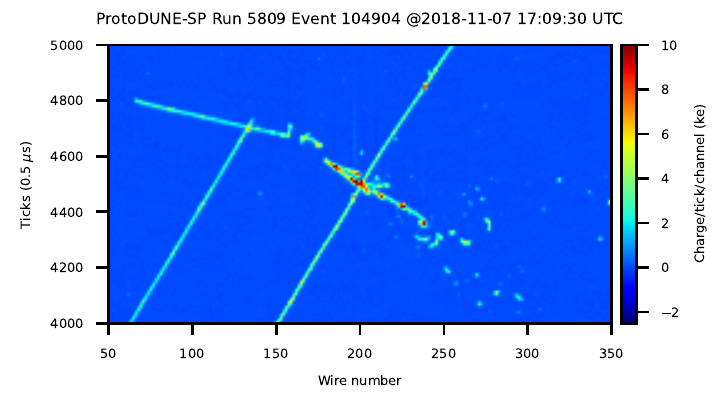}};
    \draw[-latex, thick, red] (-2.5, 1.25) to (-0.8,0.85);
    \draw[rotate around={-37:(0, 0.2)}, dashed, thick, red] (0, .2) ellipse (0.75cm and 0.3cm);
    \end{tikzpicture}
  }
  
  \subfloat[\label{fig:other_evdisp}]{
    \begin{tikzpicture}
    \node[inner sep=0pt] (image) at (0,0){\includegraphics[width=.45\textwidth]{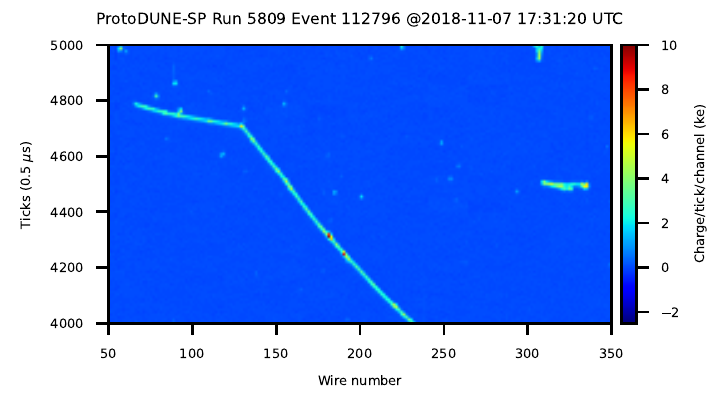}};
    \draw[-latex, thick, red] (-2.5, 1.2) to (-1.25,1.0);
    \draw[dashed, thick, red, -latex] (-1.25,1.0) to (0.85, -1.3);
    \end{tikzpicture}
  }
    \caption{Event displays of selected interaction candidates for \piplus--argon absorption~\ref{fig:abs_evdisp}, charge exchange~\ref{fig:cex_evdisp}, and other~\ref{fig:other_evdisp} interactions using the event selection described in Sec.~\ref{sec:Selection}. The candidate beam pion track begins near wire 60, tick 4800 and travels toward the bottom right of each figure, as highlighted by the solid, red arrow. The dashed ellipse in~\ref{fig:abs_evdisp} highlights two proton candidate tracks, and in~\ref{fig:cex_evdisp} it highlights a candidate $\gamma$ resulting from a $\pi^0$ decay. In~\ref{fig:other_evdisp}, the secondary $\pi^\pm$ is highlighted with a dashed arrow. Candidate cosmic-ray tracks can be seen crossing through the figures as well.}
    \label{fig:evdisps}
\end{figure}
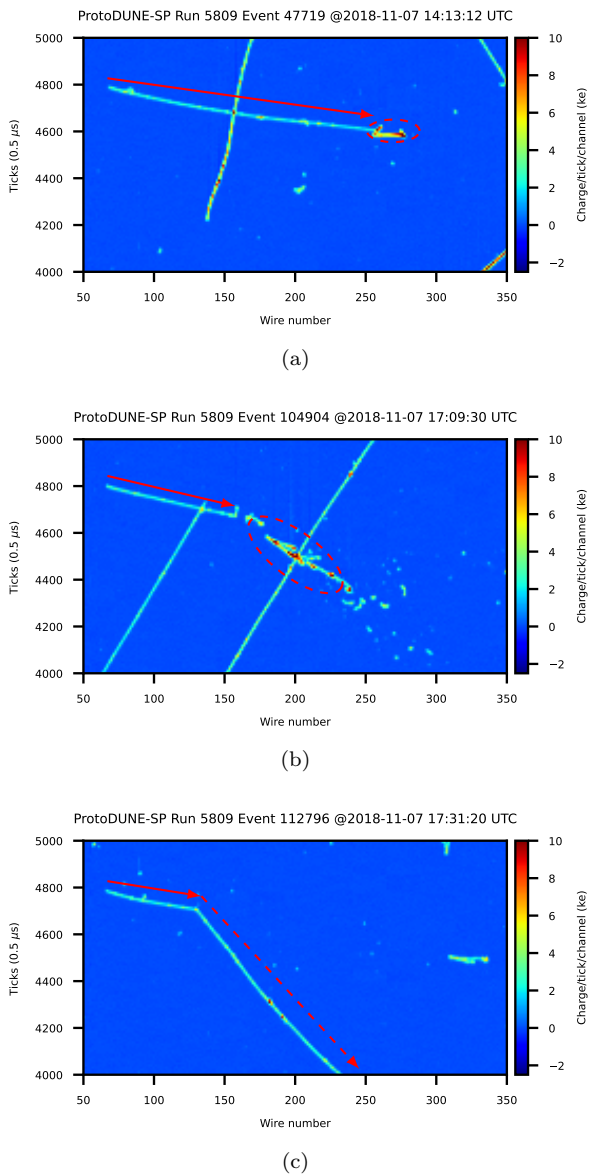

Example candidates for the three signal interactions are shown in Fig.~\ref{fig:evdisps}. The resulting purities (the number of true interaction events selected as that interaction divided by the number of true interactions of that type) and efficiencies (the number of interactions of any true category selected as a given interaction candidate divided ) within the set of events passing the pre-selection are shown in Tab.~\ref{tab:eff_purs}. The ``overall efficiency" --- in which the denominator also includes events failing the pre-selection --- is also shown.
\begin{table}[!htb]
\centering
\begin{tabular}{cccc}
\toprule
\toprule
Channel & Efficiency & Purity & Overall Efficiency\\
\midrule
Absorption & 63\% & 63\% & 36\%\\
Charge Exch. & 42\% & 75\% & 20\%\\
Other & 69\% & 63\% & 31\%\\
\bottomrule
\bottomrule
\end{tabular}
\caption{Efficiencies and purities of the event selection. Note that the first two columns are relative to the events that pass the pre-selection, while the third column is relative to the entire simulation sample.}
\label{tab:eff_purs}
\end{table}

\begin{figure*}
  \subfloat[Absorption Candidates\label{fig:abs_prefit}]{
    \includegraphics[width=.45\textwidth]{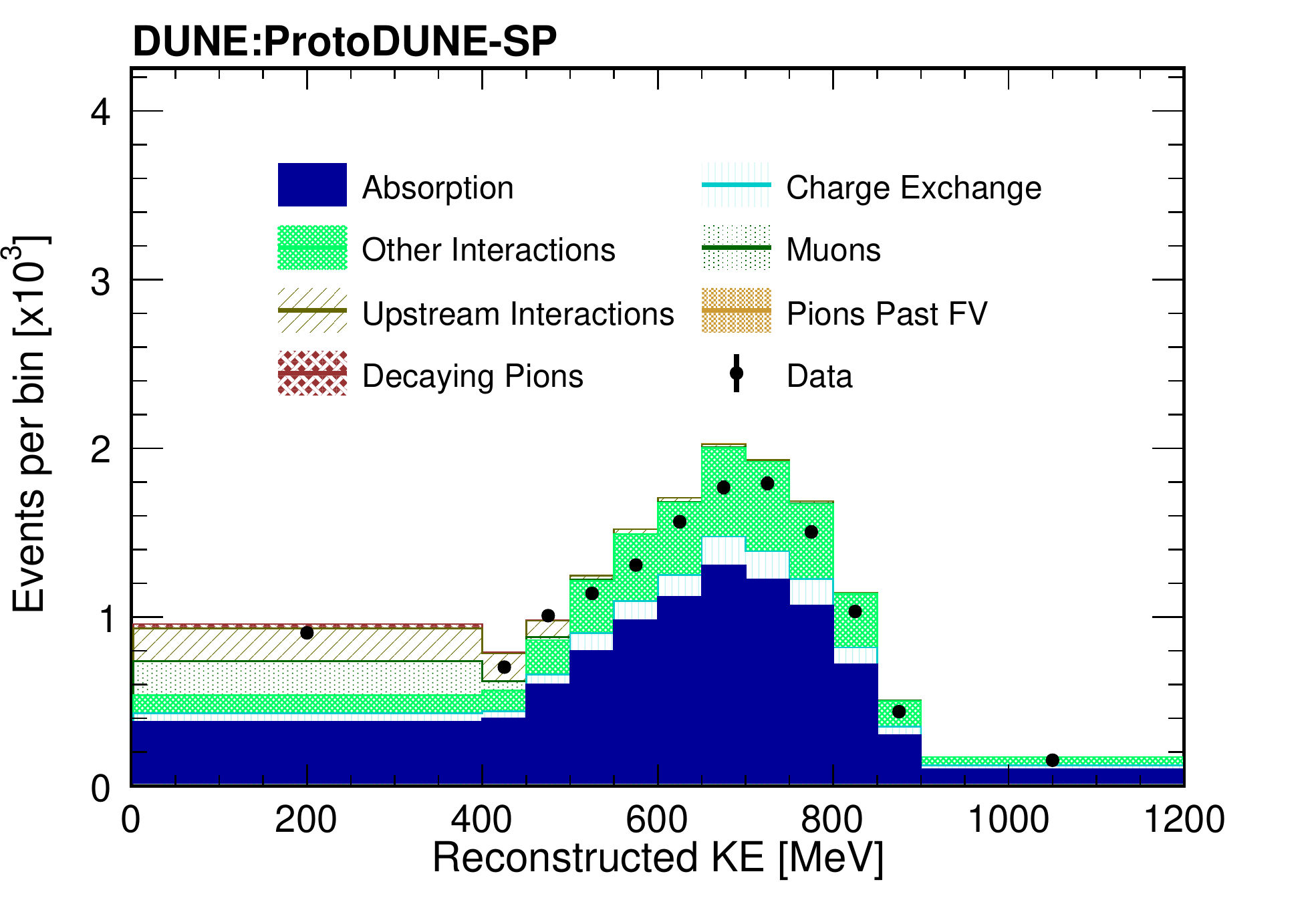}
  }
  \subfloat[Charge Exchange Candidates\label{fig:cex_prefit}]{
    \includegraphics[width=.45\textwidth]{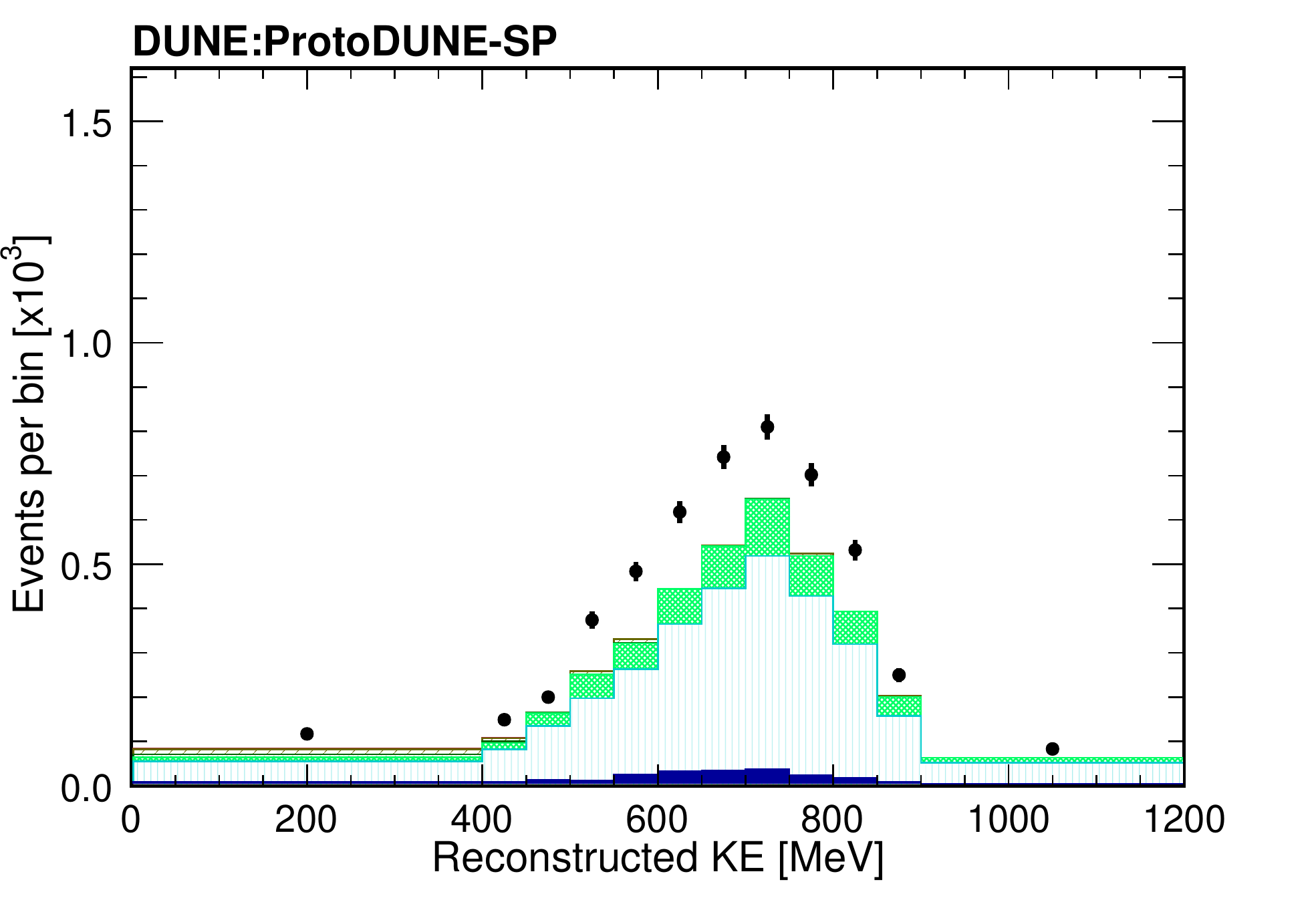}
  }
  
  \subfloat[Other Interaction Candidates\label{fig:other_prefit}]{
    \includegraphics[width=.45\textwidth]{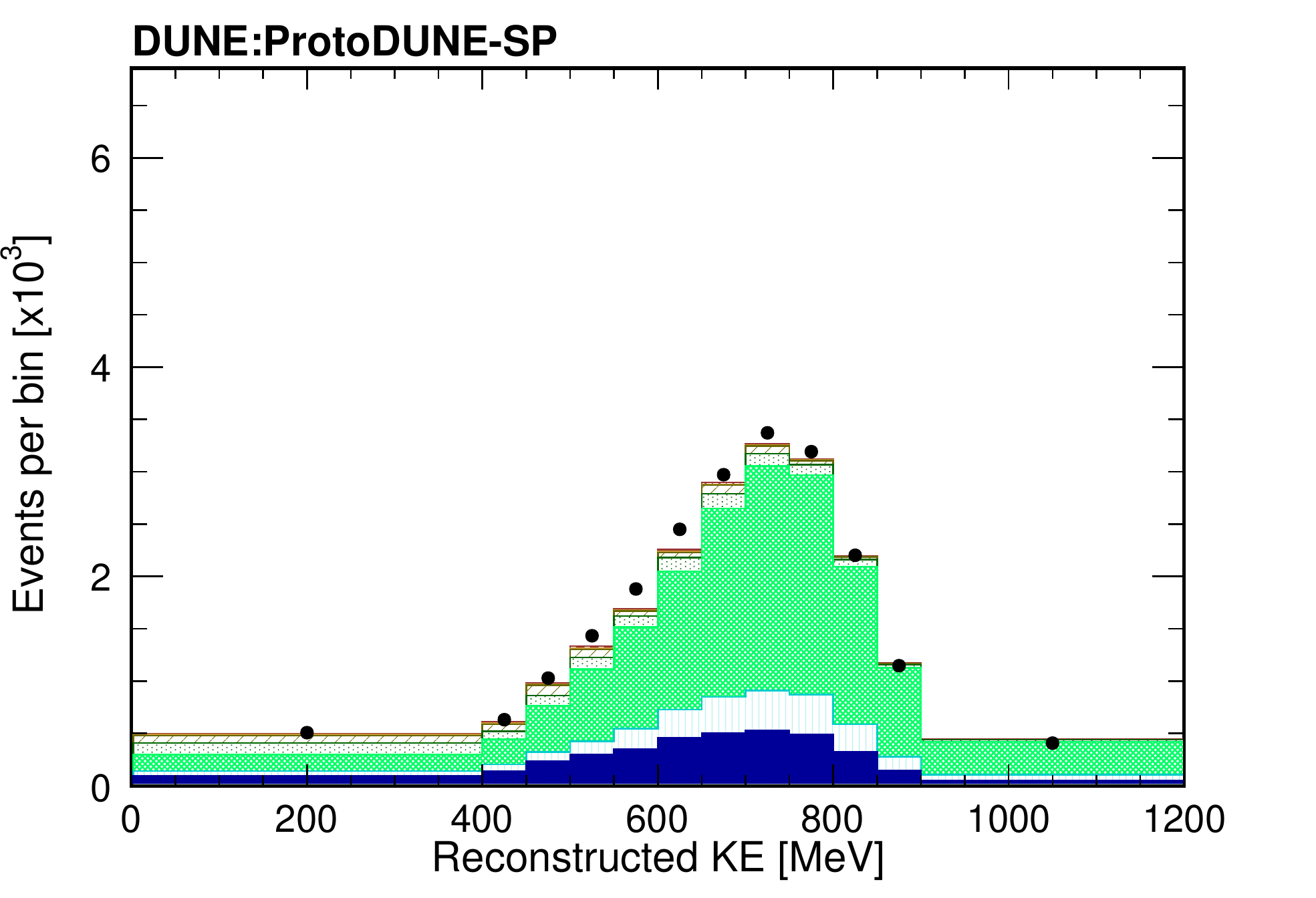}
  }
  \subfloat[Escaping Tracks\label{fig:apa2_prefit}]{
    \includegraphics[width=.45\textwidth]{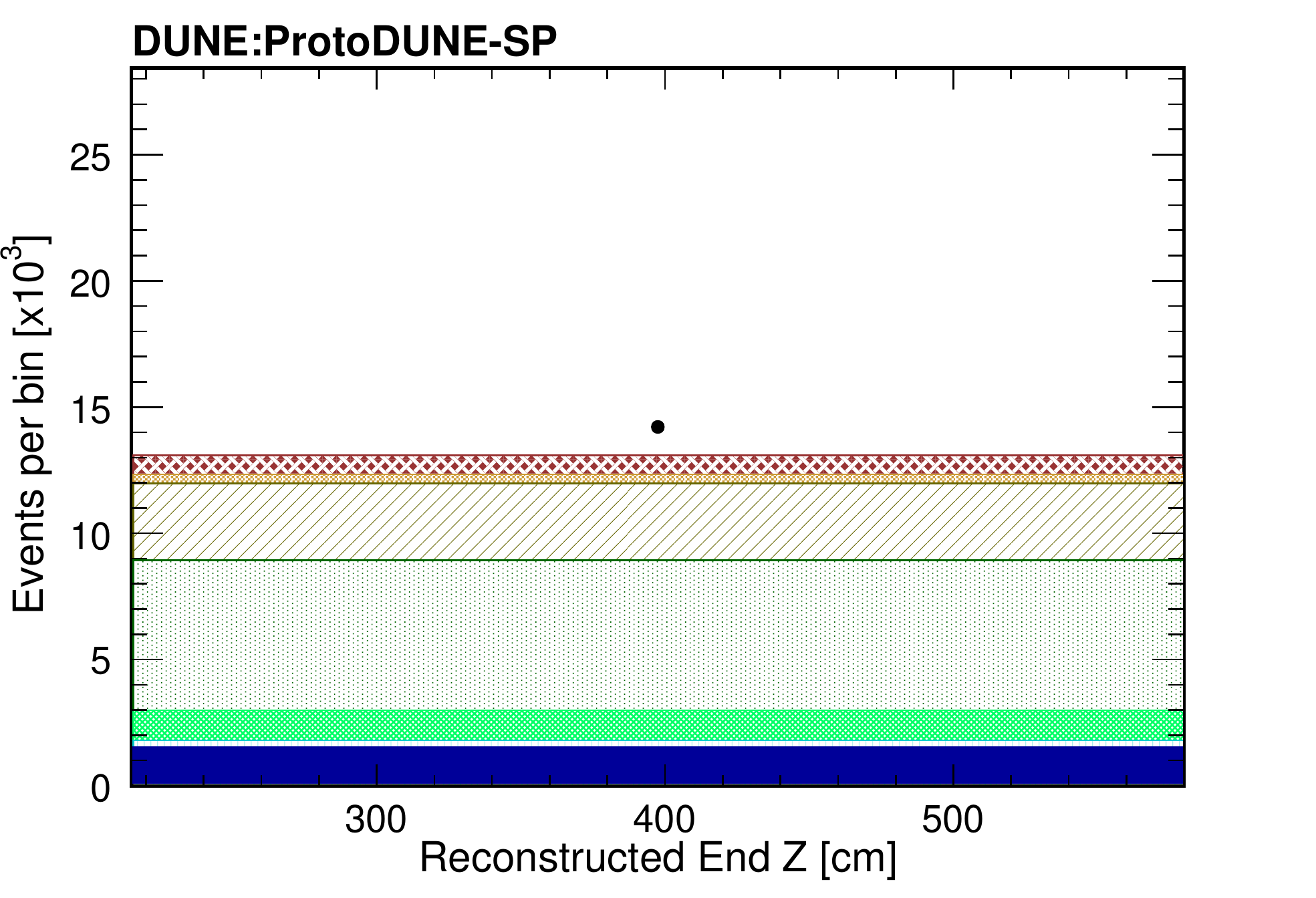}
  }

  \caption{Selected event distributions from the ProtoDUNE-SP MC with stacked contributions from true event types along with event distributions observed in data.}
  \label{fig:prefit_events}
\end{figure*}

The selected distributions resulting from the pre-fit MC (after normalizing according to the reconstructed beam line momentum distribution in data) are shown in Fig.~\ref{fig:prefit_events} and are broken down by category from MC information: the three signal interactions, beam muons, pions which interacted upstream of the TPC active volume, pions which exited the fiducial volume, and pions which stopped or decayed within the fiducial volume. Distributions collected from data are superimposed on the MC distributions.
The absorption, charge exchange, and other interactions are binned according to the reconstructed kinetic energy at the identified interaction point. This is calculated by subtracting the primary pion track's total deposited energy from the kinetic energy reconstructed within the beam line.
Events in which the reconstructed beam track extended past the limit in $z$ are contained in a single bin as shown in Fig.~\ref{fig:apa2_prefit}. We observe several differences between data and simulation in the shape and magnitude of several distributions. The observed differences are consistent with coming from two effects: 1) the number of interactions upstream of the active volume of the TPC and 2) differences in the cross sections for the signal interactions here. As will be described later, we account for the first during the fit to data using a nuisance parameter, and intend to measure the differences in physics resulting from the second point.  

\subsection{Fitting Strategy}\label{sec:llhfit}
The MC simulation is fit to the observed data in order to estimate the number of interactions as a function of the pion kinetic energy at the point of interaction. A binned likelihood fit is performed in which the fit parameters include a set of signal parameters which scale the number of signal events (absorption, charge exchange, and other inelastic interactions of primary beam \piplus) in bins of true kinetic energy at interaction. These true kinetic energy bins are 100 MeV in width and range from 500 to 800 MeV for each channel. These nine bins form the set of measurement bins within this analysis. Their widths were chosen according to the resolution of the energy reconstruction for these events. The total range (500--800 MeV) was chosen according to where the selection efficiency begins to fall off as function of kinetic energy. Scaling parameters are also applied to the underflow and overflow regions for the three channels, though the cross sections are not measured in these regions. A parameter that scales the number of interactions between the end of the beam line and the beginning of the TPC volume is included.

During each step of the fit, after the fit parameters are applied to the MC events, the MC events are reweighted according to the momentum distribution in data as reconstructed by the beam line instrumentation as described previously in Sec.~\ref{sec:Selection}. This is done such that the primary change to the simulation is the rate of interactions at various energies, rather than changing the initial momentum distribution.

The fit minimizes the negative log-likelihood ratio:
\begin{equation}
\begin{split}\label{eq:n2LL_Stat_Poisson_BB}
    &-2\textrm{log}(\lambda) = \\
    &\quad 2\sum\limits_j \bigg(\beta_j N_j^{\textrm{MC}} - N_j + N_j\textrm{log}\frac{N_j}{\beta_j N_j^{\textrm{MC}}} + \frac{(\beta_j - 1)^2}{2\sigma_j^2}\bigg)
\end{split}
\end{equation}
Eq.~\ref{eq:n2LL_Stat_Poisson_BB} is derived from the Poisson likelihood to observe $N_j$ events in the set of reconstructed bins $j$ in which $\beta_jN_j^{\textrm{MC}}$ events are predicted. $\beta_j$ are scaling factors proposed by Barlow and Beeston which account for MC statistical uncertainty and reflect the fact that the generated MC events $N_j^{\textrm{MC}}$ are themselves drawn from some underlying distribution with mean $\beta_jN_j^{\textrm{MC}}$~\cite{barlowbeeston}. Barlow and Beeston treat these factors as additional nuisance parameters in the fit. However, we follow Conway's treatment~\cite{Conway}, in which a Gaussian constraint\footnote{With variance $\sigma^2_j = \sum\limits_i w^2_i / (\sum\limits_i w_i)^2$. $w_i$ is the weight of event $i$ in bin $j$. The sums run over each event in bin $j$.} is added to the likelihood and the factors are determined analytically by setting the derivative of the likelihood with respect to each $\beta_j$ to 0 (maximizing the likelihood).


\subsection{Iterative Fit Procedure}
In order to mitigate any dependence of the analysis results on the choice of signal model, we developed an iterative fit procedure. In each subsequent fit, the signal model is updated using Geant4Reweight~\cite{geant4reweight} which produces event weights to effectively modify the \piplus~cross sections in the input MC to match the cross sections extracted from the previous fit.

Tests were performed using fake data produced from MC varied by Geant4Reweight in order to verify the convergence of this technique. This procedure reduces the bias to less than 1\% of the true cross sections. This residual bias is negligible compared to the systematic uncertainties. Additionally, a data-driven stopping criterion was developed as follows. For a given set of data (fake or real), we fit to 500 statistically-varied distributions drawn from the data set. For each of these fits, the result is used to update the model, and the fit is performed again using the update.  We calculate the $\chi^2$ between each updated fit and its previous iteration, using the distribution of extracted cross sections values from the updated iteration to calculate the covariance used within the $\chi^2$ calculation. This $\chi^2$ represents how much the model continues to change in each subsequent fit. We continue updating and refitting until $90\%$  of fits end with $\chi^2 < 1$ indicating that the model has stopped updating. In the case of this measurement, we stopped iterating at the fourth fit. The differences between the cross sections from the first and fourth fit to data relative to the fourth-fit cross sections are shown in Tab.~\ref{tab:biases}.

\begin{table*}[!htb]
\centering
\begin{tabular}{c|ccc}
\hline\hline
\multirow{2}{*}{$T_{\pi} \textrm{[MeV]}$} & \multicolumn{3}{c}{Relative Difference} \\
\cline{2-4}\\[-0.75em]
& Absorption & Charge Exchange & Other Interactions \\
\hline
550 & $-7.1\%$ & $+11.2\%$ & $+8.3\%$\\
650 & $-3.8\%$ & $+5.0\%$ & $+5.2\%$\\
750 & $-4.6\%$ & $+2.6\%$ & $+3.7\%$\\
\hline\hline
\end{tabular}
\caption{Differences between the cross sections from the first and fourth iterations of the fit to data relative to the cross section from the fourth fit iteration for each $\pi^+$ kinetic energy bin.}
\label{tab:biases}
\end{table*}

\section{Treatment of Uncertainties}\label{sec:uncertainties}
In order to propagate uncertainties within the analysis, alternate fits are performed by varying an aspect of either some aspect of the simulation or, in one case, the calibration in data for each uncertainty source. These alternate cross sections are used to calculate covariances which are added in quadrature to give the full covariance of the measured cross sections.

For the statistical uncertainty associated with the collected data, alternate data sets are generated by setting the number of events in each bin of data to Poisson-distributed random integers drawn from the content of each bin . The simulation is then fit to these alternate data sets, and the contribution to the covariance of bins $i$ and $j$ is calculated as

\begin{equation}\label{eq:prop_cov}
    V_{ij} = \frac{1}{N_{\textrm{Fits}}}\sum\limits_k^{N_{\textrm{Fits}}}(\sigma_{i} - \sigma_i^k)(\sigma_{j} - \sigma_j^k).
\end{equation}
Here, $\sigma_{i(j)}$ is cross section in true bin $i$ $(j)$ from the ``standard'' (simulation and collected data) fit result and $\sigma^k_{i(j)}$ is the cross section in true bin $i$ $(j)$ from the $k^{th}$ alternate fit. We performed fits to 500 statistically varied distributions.

A similar procedure is performed for the statistical uncertainty associated with the input simulation. However, instead of changing the data distributions, the simulation distributions were varied by producing event weights calculated by drawing Poisson-distributed numbers around the number of events in the initial simulation distributions. The covariances are calculated again using Eq.~\ref{eq:prop_cov}, and we used the same number of varied fits for the calculation. The number of events from our simulation is of the same order as those in the data; the statistical uncertainties are thus very similar.
Future analyses will make use of a larger sample of input simulation to reduce this uncertainty.

\begin{figure*}[!htb]
  
  

  \subfloat[Absorption\label{fig:abs_errs}]{
    \includegraphics[width=.45\textwidth]{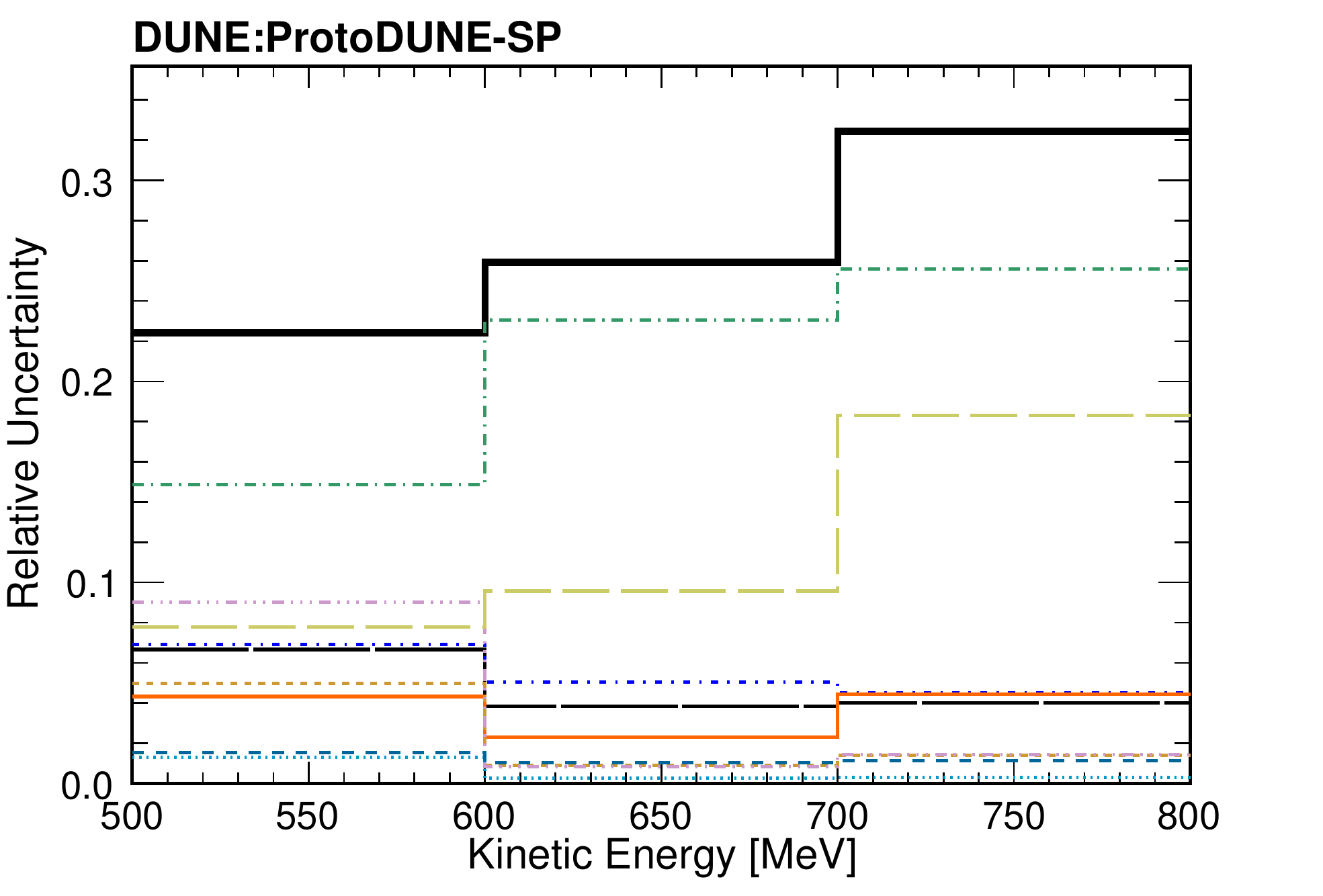}
  }
  \subfloat[Charge Exchange\label{fig:cex_errs}]{
    \includegraphics[width=.45\textwidth]{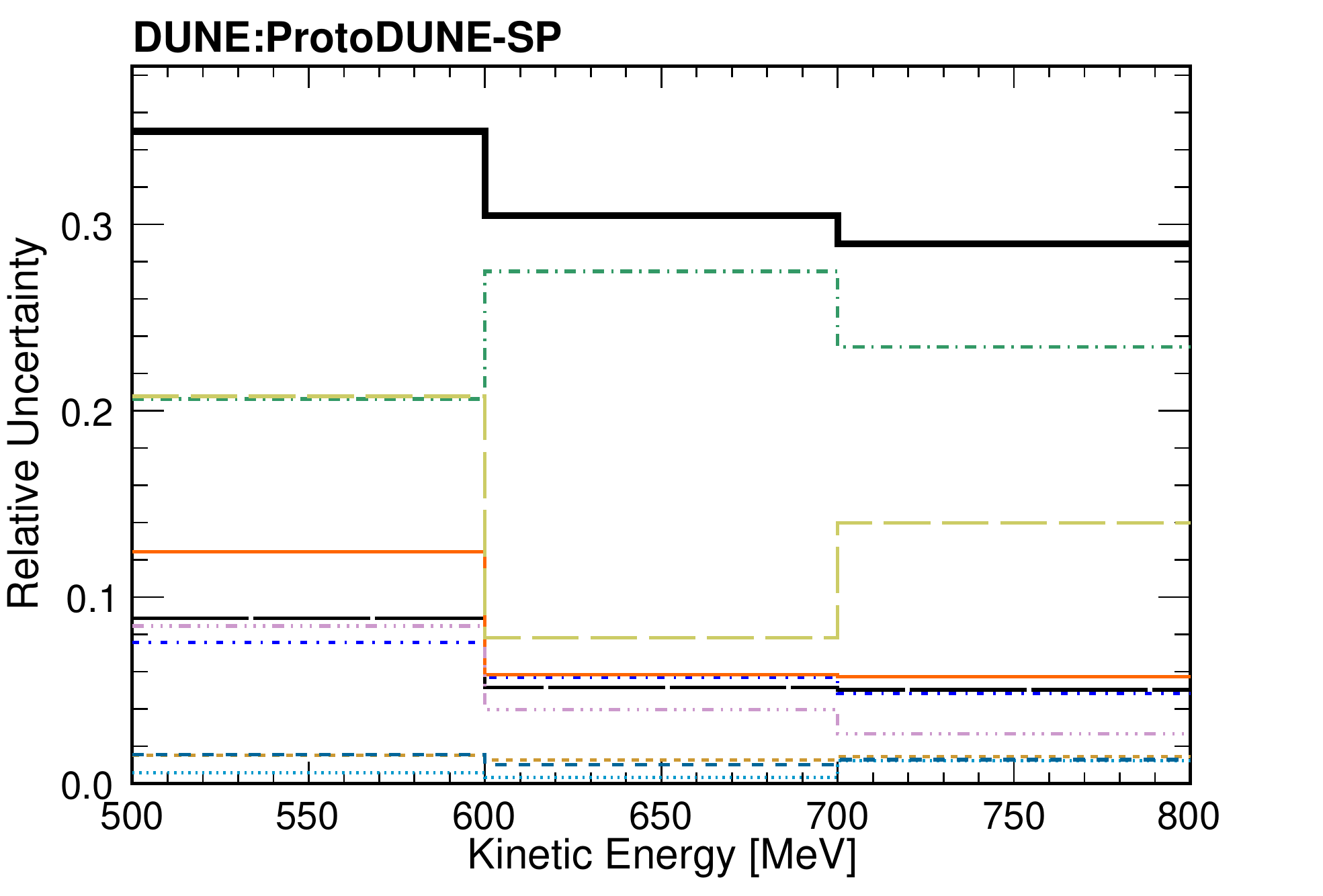}
  }
  
  \subfloat[Other Interactions\label{fig:other_errs}]{
    \includegraphics[width=.45\textwidth]{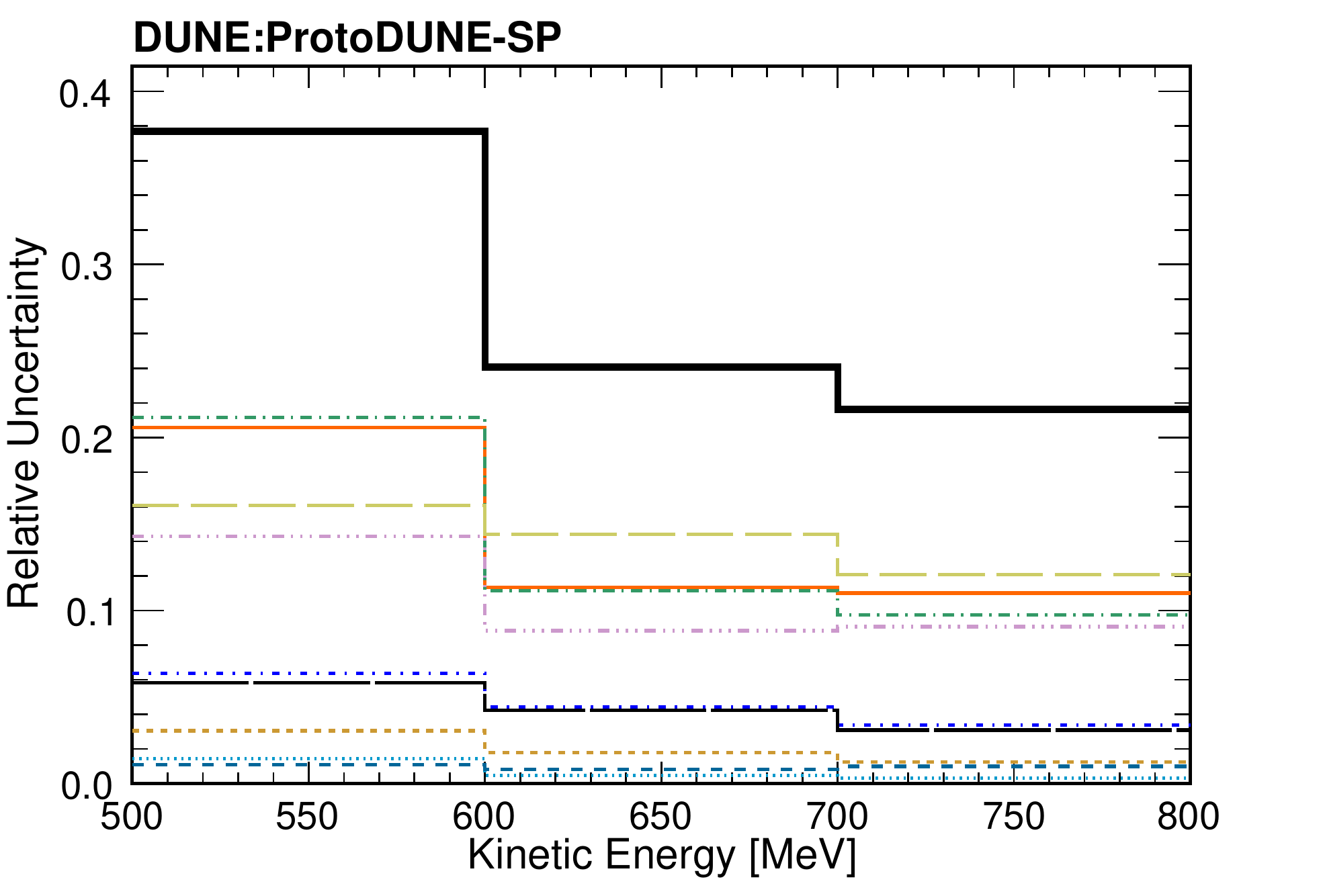}
  }
  \subfloat[\label{fig:leg_errs}]{
    \includegraphics[width=.45\textwidth]{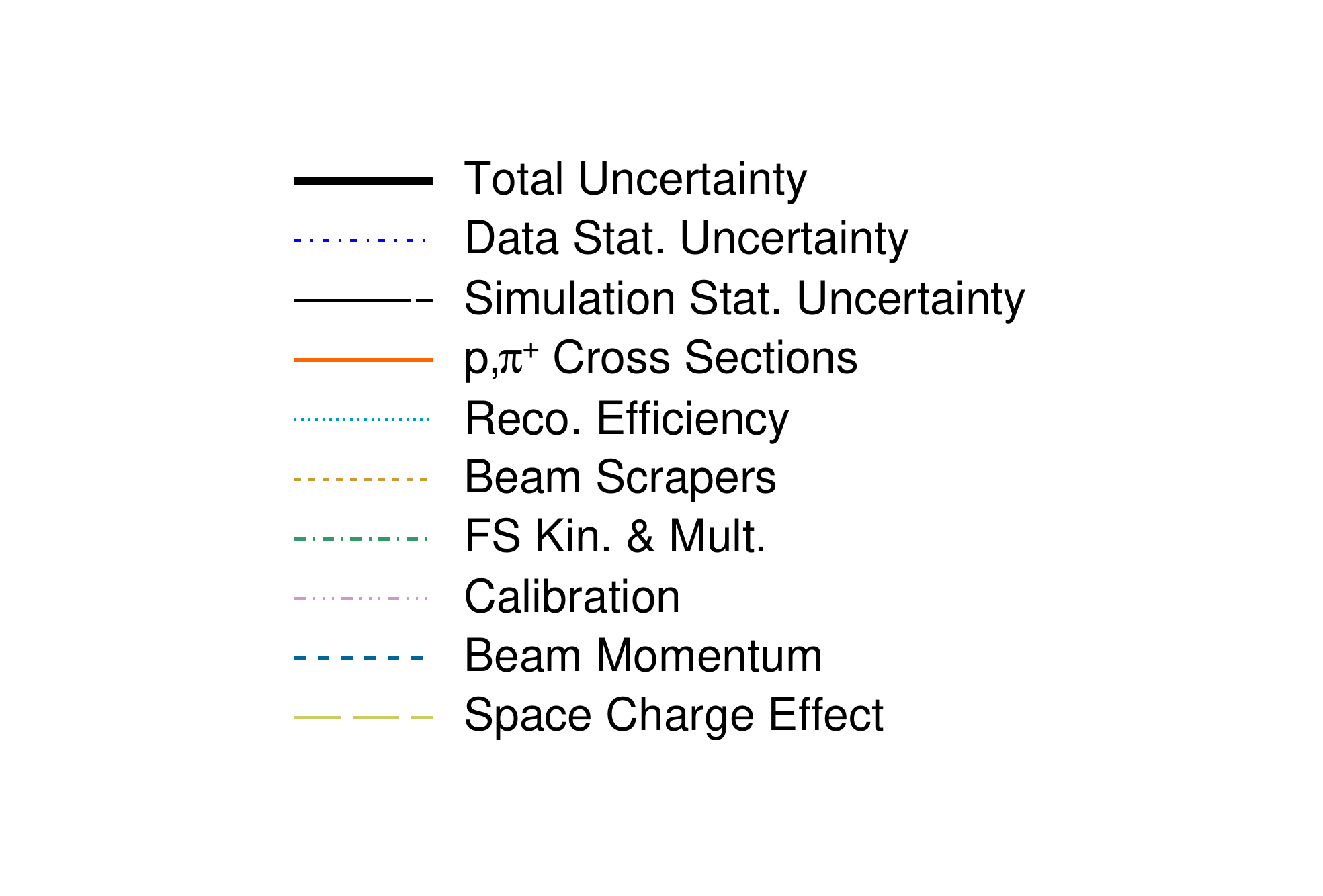}
  }

  \caption{Uncertainties relative to the best-fit cross section values for absorption~\ref{fig:abs_errs}, charge exchange~\ref{fig:cex_errs}, and other interactions~\ref{fig:other_errs}. The legend is shown in~\ref{fig:leg_errs}.}
  \label{fig:xsec_errors}
\end{figure*}

The individual systematic uncertainties are described below, and their contribution to the uncertainty of the measurement is shown in Fig.~\ref{fig:xsec_errors}. The label by which the systematic effect is referred  to within this figure is stated at the end of each item. Unless otherwise stated, their contributing covariances in bin ($i$,$j$) are calculated from fits using simulation varied with $\pm1\sigma$ shifts to the underlying effect and are given by:
\begin{equation}\label{eq:pm1sigma_cov}
\begin{split}
    V_{ij} = \frac{1}{2}\big(&(\sigma_i - \sigma^{+}_i)(\sigma_j - \sigma^{+}_j) \\\space+ &(\sigma_i - \sigma^{-}_i)(\sigma_j - \sigma^{-}_j)\big),
\end{split}
\end{equation}
where $\sigma^{\pm}$ represents the cross section extracted from the fit with the systematic effect varied by $\pm1\sigma$. Several of the systematic effects described below are broken up into multiple independent contributions to the total uncertainty. In other words, when specified, several independent variations to the fit --- and thus separate covariances calculated as in Eq.~\ref{eq:pm1sigma_cov} --- have been made. In these cases, the corresponding uncertainty contributions in Fig.~\ref{fig:xsec_errors} include these separate covariances added in quadrature.
\begin{enumerate}
    \item Beam Scrapers: the number of beam \piplus/$\mu^+$ that scrape along the beam plug between the end of the beam line and the beginning of the active TPC volume. The beam plug displaces argon along the nominal path of the beam particle so less energy is lost and fewer hadronic interactions occur before the beam particle enters the active volume of the TPC. A difference in the number of particles which scrape the plug can affect the relationship between true and reconstructed \piplus~energy at interaction. Simulation showed that the distributions in momentum of the beam \piplus/$\mu^+$ differ due to the fact that some of the muons originate from decaying pions. The effect was parameterized separately for pions across their momentum range and for two regions in the momentum range of beam muons (corresponding to different populations of muons which were produced from hadron decays at different points along the beam line). A 50\% uncertainty to the number of scraping particles in each of these three populations was estimated.
    
    \item $\textrm{p},\pi^+$ Cross Sections: the total interaction rate of $\pi^+$ and protons. A misestimation of the inelastic cross sections can affect selection efficiencies and purity. As such, event weights were produced using Geant4Reweight~\cite{geant4reweight} to represent variations to the proton and \piplus cross sections. One parameter was used to vary the proton cross section across a momentum range encompassing all protons produced within the MC sample. Two parameters were used for \piplus~interactions: 1) below 600 MeV/$c$ and 2) above 600 MeV/$c$ pion momentum. A 33\% uncertainty was assigned to the cross section parameters and was estimated from the spread of various hadron--nucleus interaction models. The separation of the systematic effect for \piplus into two momentum regions was informed by the varying final state behavior between the models~\cite{NEUTTune}.

    \item FS Kin. \& Mult.: final state kinematics \& multiplicities of the primary interaction. The selection efficiency can depend on the distribution in various dimensions of phase space of final state particles emitted from the primary interactions. By comparing to either GENIE's \hA/\hN~ models or \geant's INCL++ hadronic interaction model, two alternate simulation sets were produced for each of the following projections: 
    \begin{itemize}
        \item Momentum of the leading outgoing $p$
        \item Momentum of the leading outgoing \piplus
        \item Momentum of the leading outgoing \pizero
        \item Angle (relative to the primary \piplus) of the leading $p$
        \item Angle of the leading \piplus
        \item Angle of the leading \pizero
        \item $p$ multiplicity
        \item $n$ multiplicity
        \item \piplus~multiplicity
        \item \piminus~multiplicity        
        \item \pizero~multiplicity
    \end{itemize}
    The corresponding uncertainty for this effect as shown in Fig.~\ref{fig:xsec_errors} includes all of the projections above added together in quadrature.
    
    \item Reco. Efficiency: beam reconstruction and matching efficiencies. Systematic effects arise in reconstruction due to differences between data and MC in the incident beam's direction and position on the face of the detector. Two parameters were implemented to account for these effects. The first varies the efficiency of reconstructing tracks to beam particles with short true trajectory length, while the second varies the rate of Pandora's ability to correctly choose the true test beam particle itself as the reconstructed beam rather than choosing a cosmic ray particle as the beam. For the former, we performed a study using selected beam tracks, wherein the tracks were artificially shortened by removing reconstructed hits from the event record until the track was a certain length. From this, we observed a 26$\%$ difference between data and MC performance in the ability to correctly reconstruct these shortened tracks, and we used this value as the uncertainty on the effect. For the second effect, we used a conservative 100$\%$ relative uncertainty on the rate of misidentification. 
    
    \item Beam Momentum: differences in the relationship between momentum reconstructed by the spectrometer in the beam line and the true momentum at the beginning of the TPC active volume can lead to a different smearing between true and reconstructed energy at \piplus~interaction --- one of the primary observables in the fit. This relationship was found to be described by a second-order polynomial, and a sample of stopping beam particles was used to determine the covariance between this polynomial's coefficients. 500 alternate simulation sets were produced by sampling these coefficients. These alternate sets were fit to data, and the covariance was calculated as in Eq.~\ref{eq:prop_cov}. 
    
    \item Calorimetry Calibration: An integral part of the calibration is determining the overall charge scale of the detector. This is used in converting a track's $dQ/dX$ (charge reaching the readout wires per unit length) into its $dE/dX$ (energy deposited per unit length). A 3\% uncertainty was assigned to the overall charge scale factor used within the $dQ/dX$ to $dE/dX$ conversion (a slightly more conservative value according to the study performed in Ref.~\cite{PDSPResults}). 
    
    \item Space Charge Effect: as in Ref.~\cite{KaonMeasurement}, studies using a sample of cosmic ray tracks found an 8\% uncertainty on the magnitude of spatial distortions arising due to SCE. Alternate MC samples were produced where the distortions were increased/decreased by 8\% and used to fit the standard data set. 

\end{enumerate}

\begin{table*}[!htb]
\centering
\begin{tabular}{cccc|c}
\hline\hline
$T_{\pi} \textrm{[MeV]}$ & $\sigma_{\textrm{Abs}} [\textrm{mb}]$ & $\sigma_{\textrm{Cex}} [\textrm{mb}]$ & $\sigma_{\textrm{Other}} [\textrm{mb}]$  & $\sigma_{\textrm{Tot}} [\textrm{mb}]$\\
\hline
550 & 160 $\pm$ 36 & 148 $\pm$ 52 & 276 $\pm$ 104 & 584 $\pm$ 122\\
650 & 161 $\pm$ 42 & 144 $\pm$ 44 & 253 $\pm$ 61  & 558 $\pm$ 86\\
750 & 131 $\pm$ 43 & 158 $\pm$ 46 & 301 $\pm$ 65  & 590 $\pm$ 90\\
\hline\hline
\end{tabular}
\caption{Summary of cross section results for each $\pi^+$ kinetic energy bin. The second, third, and fourth columns are the three exclusive signal channels. The far right column is the total inelastic cross section.}
\label{tab:xsecs}
\end{table*}

\section{Results}\label{sec:results}
\begin{figure*}[!htb]
  \subfloat[Absorption Candidates\label{fig:abs_events}]{
    \includegraphics[width=.45\textwidth]{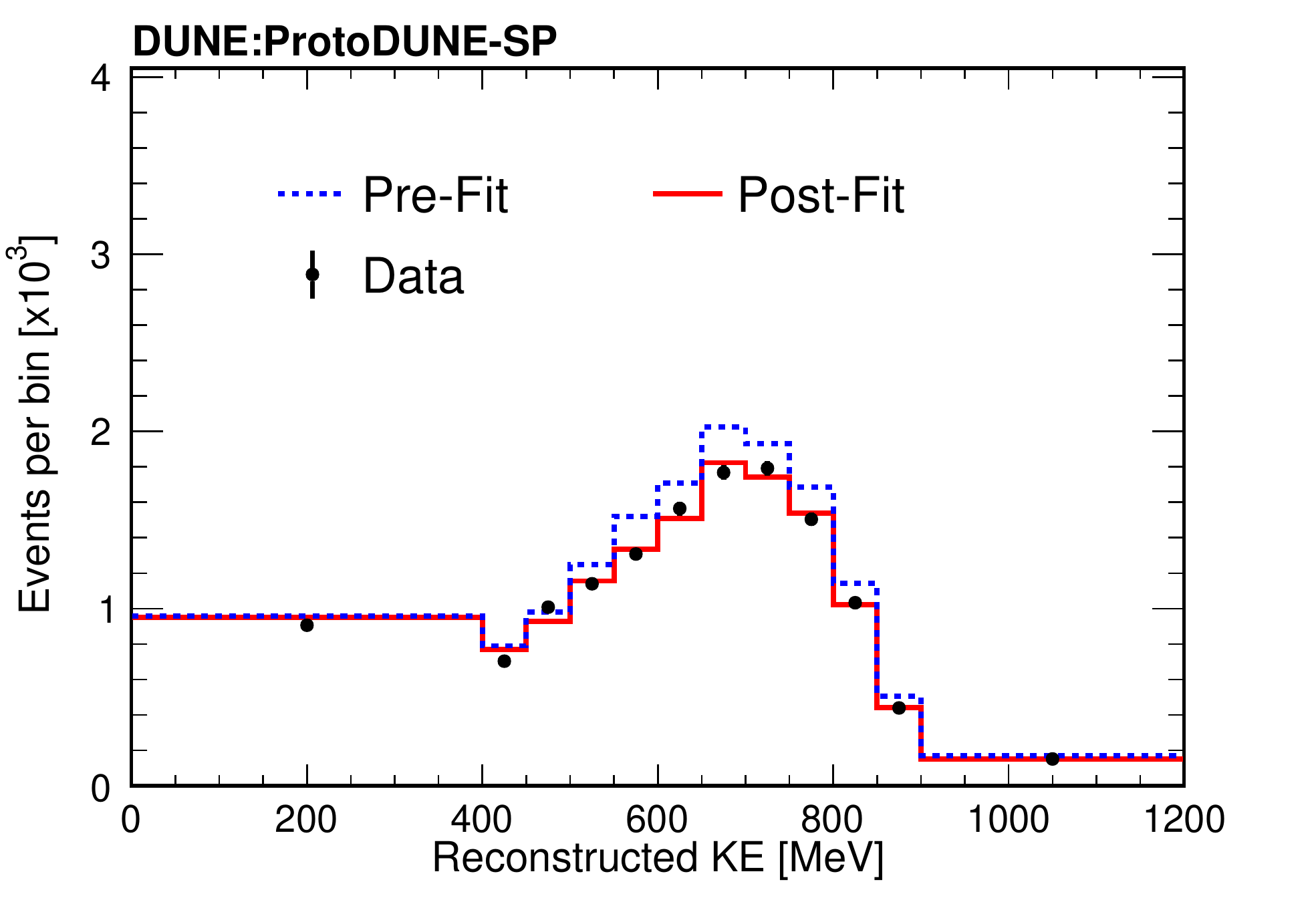}
  }
  \subfloat[Charge Exchange Candidates\label{fig:cex_events}]{
    \includegraphics[width=.45\textwidth]{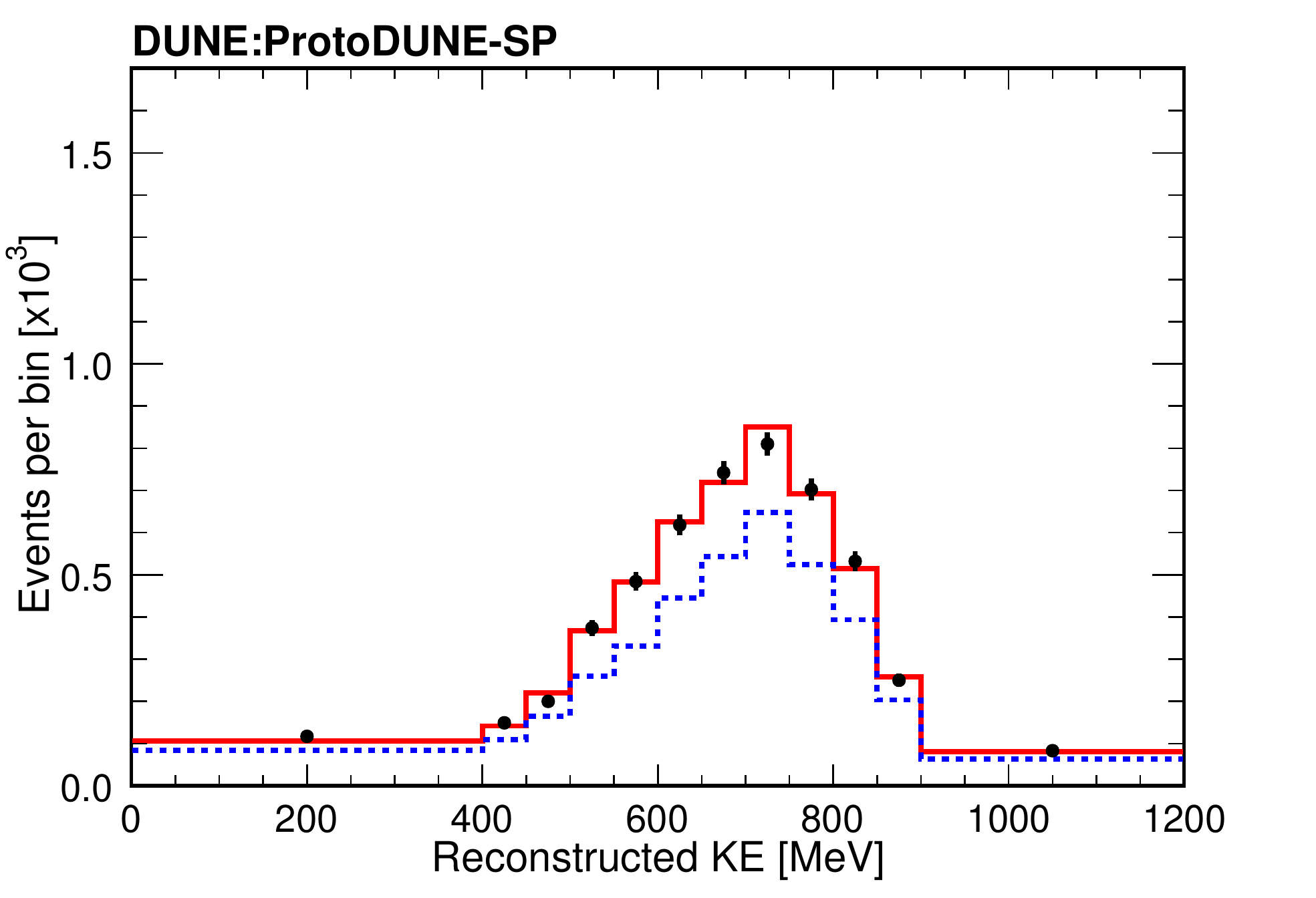}
  }
  
  \subfloat[Other Interaction Candidates\label{fig:other_events}]{
    \includegraphics[width=.45\textwidth]{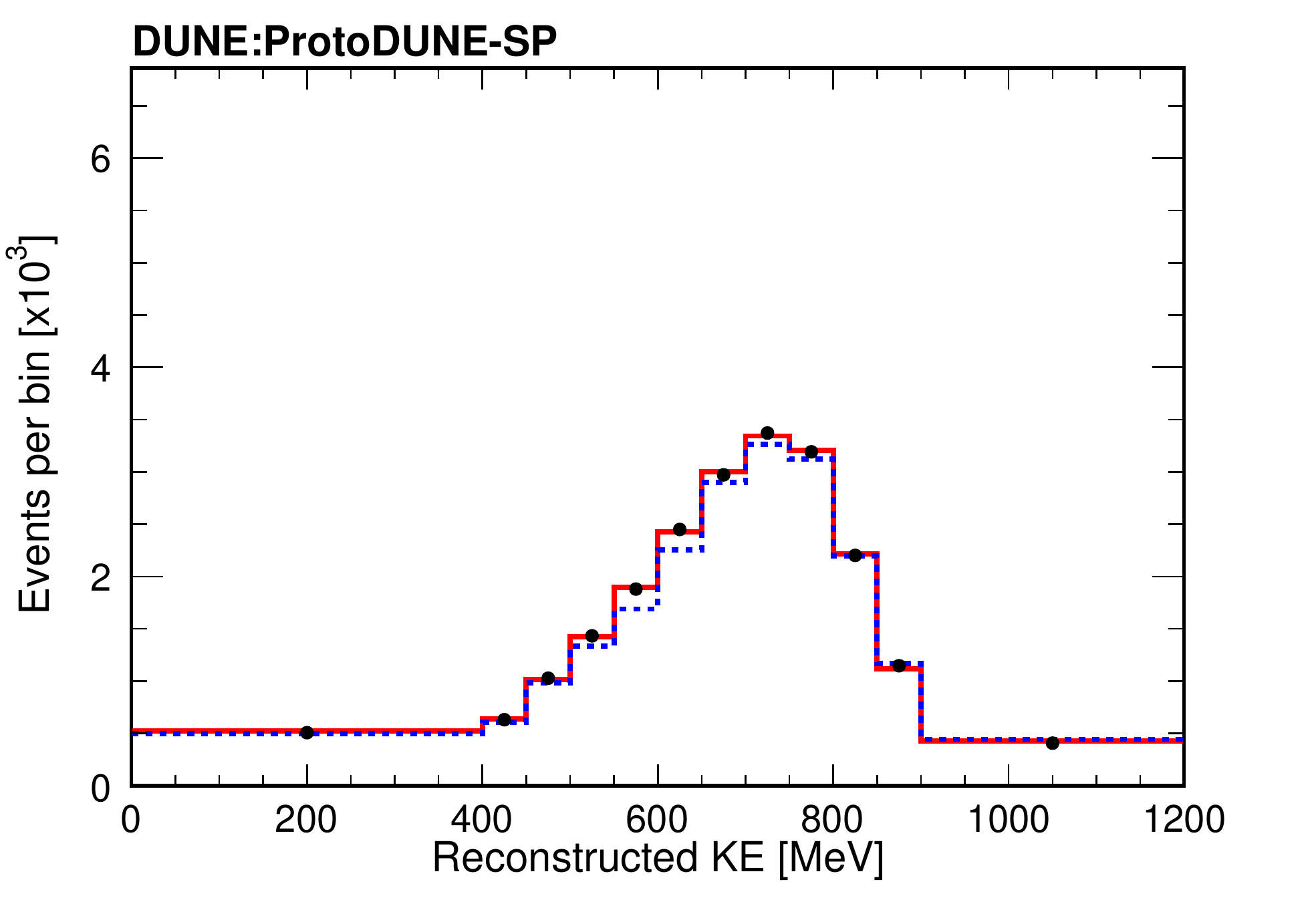}
  }
  \subfloat[Escaping Tracks\label{fig:apa2_events}]{
    \includegraphics[width=.45\textwidth]{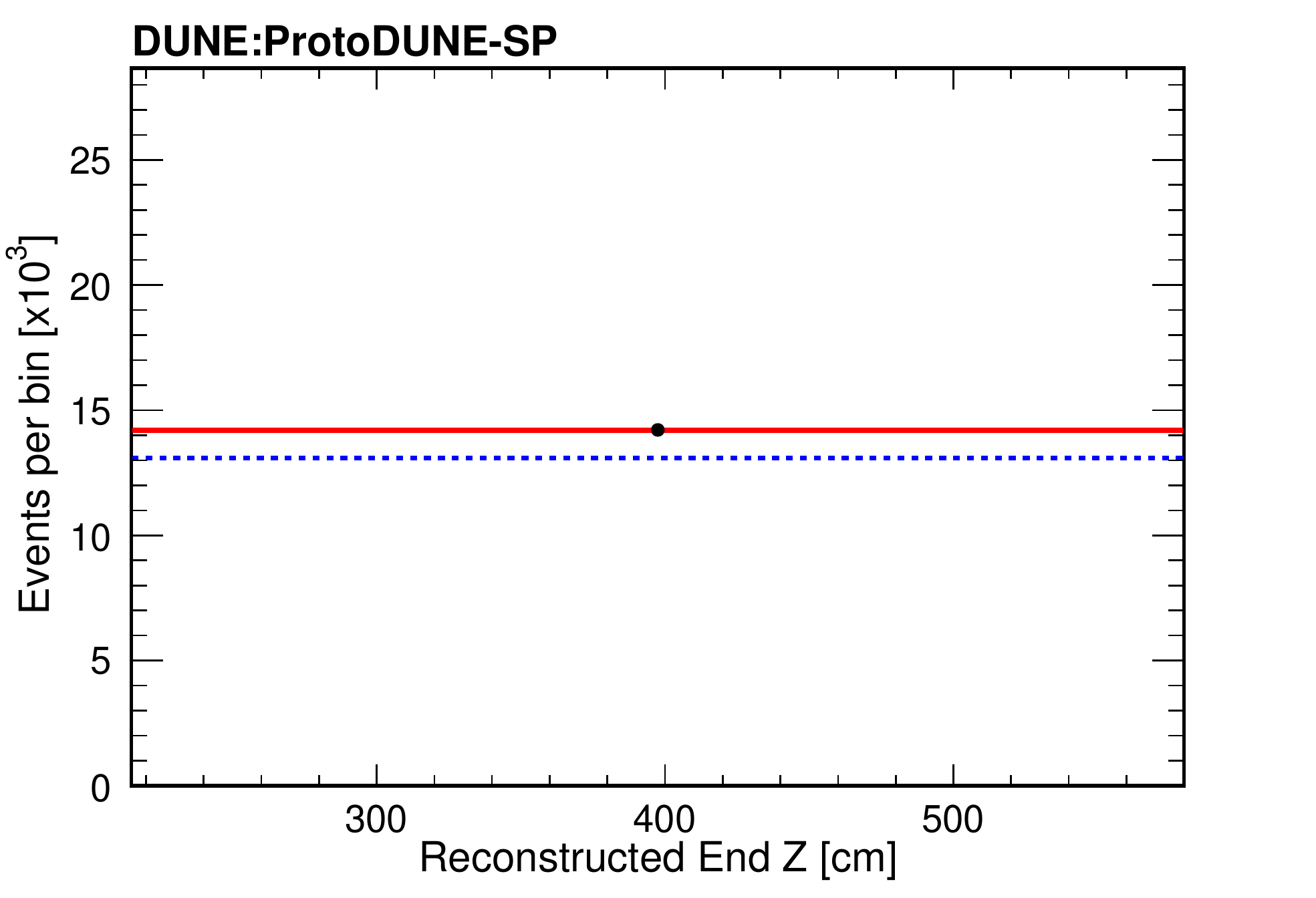}
  }
  
  \caption{Event distributions observed by ProtoDUNE-SP (black circles) compared to pre-fit (dashed blue histogram) and post-fit MC (solid red histogram) using 4 fit iterations. The error bars on the bins of the data distribution (often smaller than the data point markers themselves) represent the estimated statistical errors.}
  \label{fig:results_events}
\end{figure*}

Fig.~\ref{fig:results_events} shows reconstructed event distributions collected in data and as predicted by the pre-fit and post-fit MC (for which our data-driven iterative extraction procedure ran for 4 iterations) in the various samples described in Sec. \ref{sec:Selection}.  The distributions show good agreement between the post-fit MC and data.

\begin{figure*}[!htb]
  \subfloat[Absorption\label{fig:abs_xsec}]{
    \includegraphics[width=.32\textwidth]{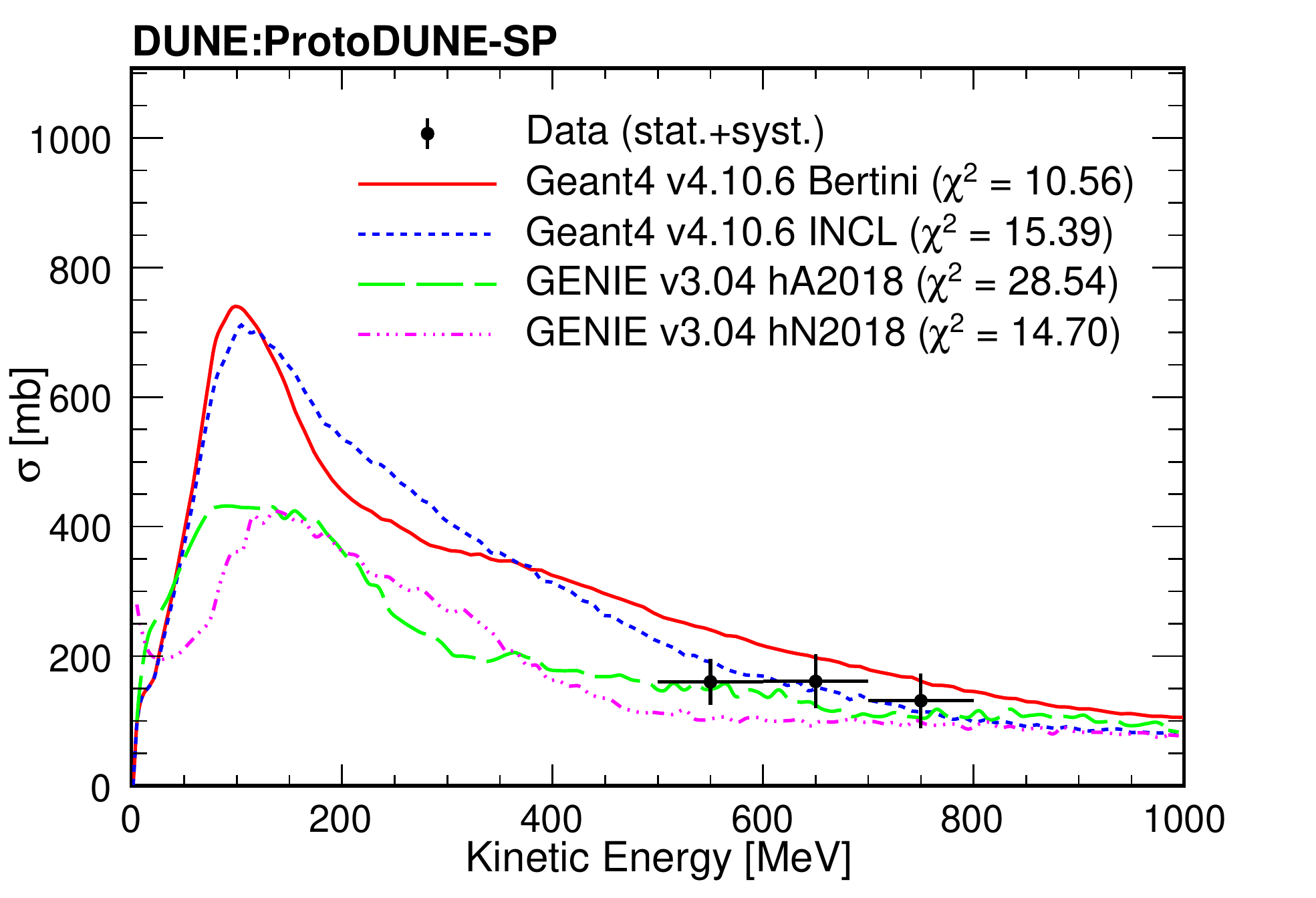}
  }
  \subfloat[Charge Exchange\label{fig:cex_xsec}]{
    \includegraphics[width=.32\textwidth]{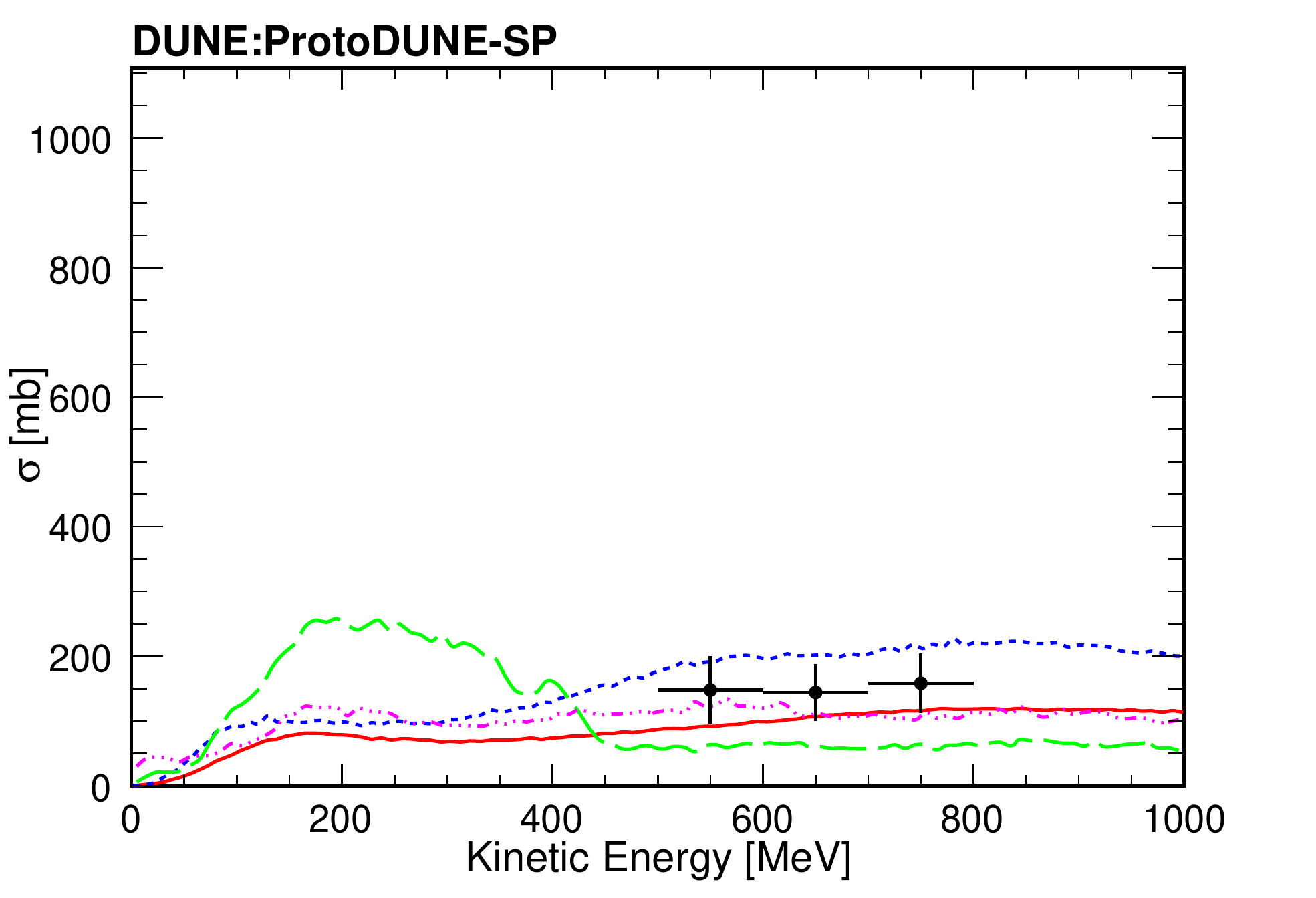}
  }
  \subfloat[Other Interactions\label{fig:other_xsec}]{
    \includegraphics[width=.32\textwidth]{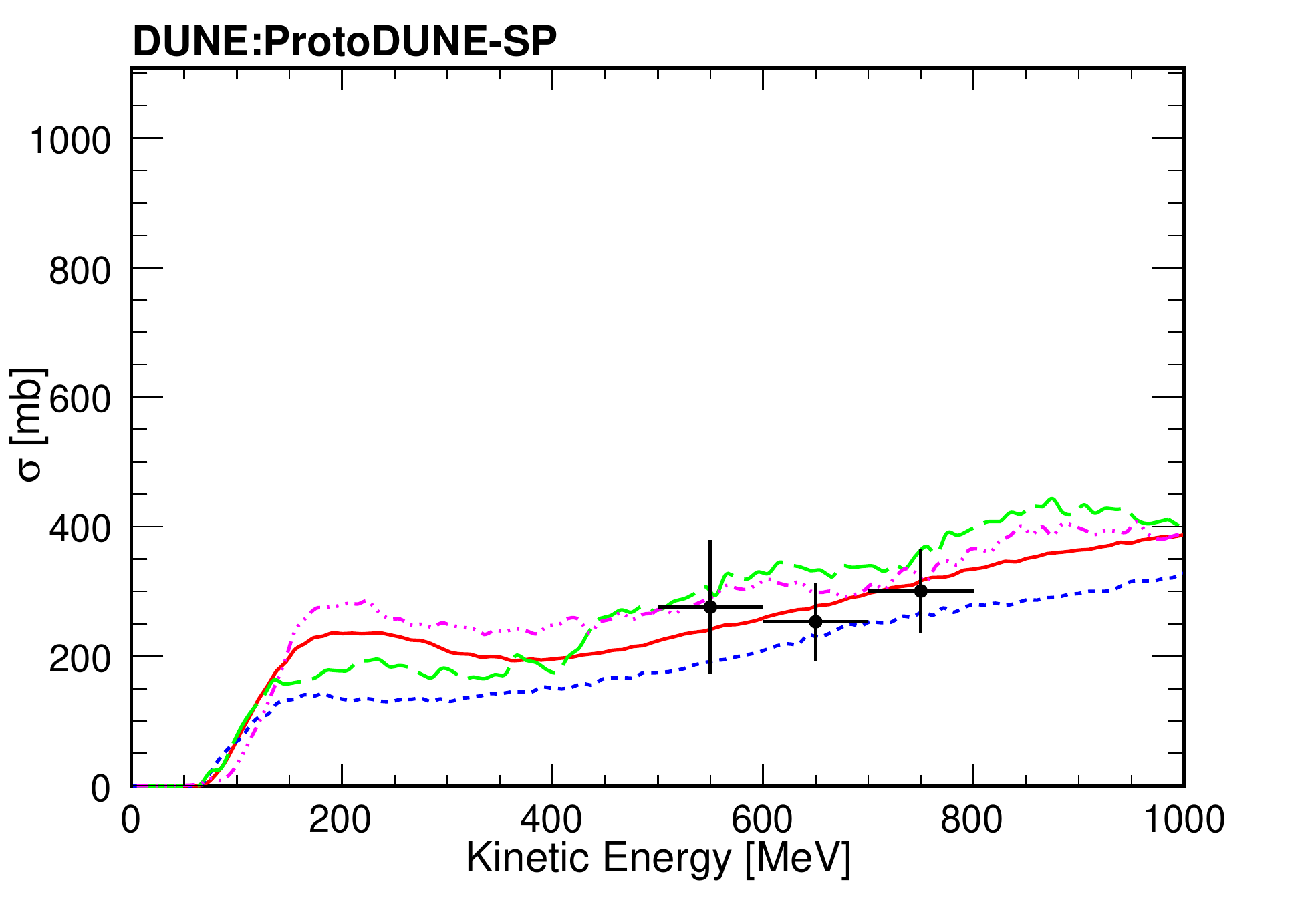}
  }
  
  \subfloat[Correlation Matrix\label{fig:corr_xsec}]{
    \includegraphics[width=.32\textwidth]{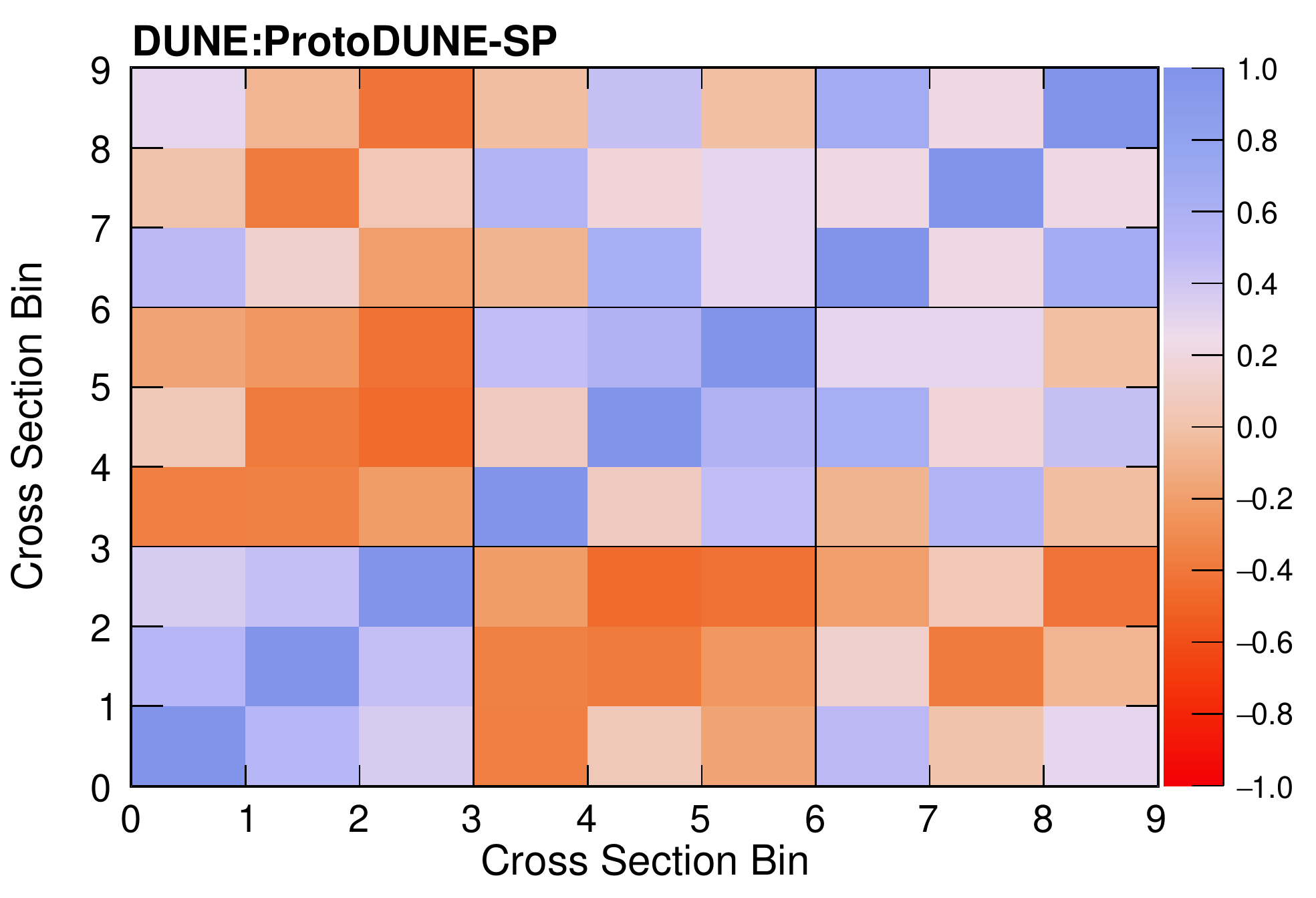}
  }

  \caption{ProtoDUNE-SP measurement of the cross sections for absorption~\ref{fig:abs_xsec}, charge exchange~\ref{fig:cex_xsec}, and other interactions~\ref{fig:other_xsec} compared to \geant~and \genie~predictions. The error bars represent the combined statistical and systematic uncertainties (added in quadrature). \ref{fig:corr_xsec} shows the correlations between the measured cross sections. The bins are arranged as: absorption (1, 2 3), charge exchange (4, 5, 6), and other (7, 8, 9) in order of increasing kinetic energy. The black lines are guides to highlight the sections of the correlation matrix representing different exclusive channels. The $\chi^2$ shown in~\ref{fig:abs_xsec} are calculated using the nine bins across the three channels and their covariances.}
  \label{fig:results_xsecs}
\end{figure*}

Fig.~\ref{fig:results_xsecs} shows the exclusive cross sections measured within this analysis in comparison to those predicted by the Bertini and INCL cascade models within \geant~\cite{geant4phys} as well as the \genie~FSI models $hA$ and \hN~\cite{genie}. The model predictions shown in Fig.~\ref{fig:results_xsecs} include the signal definition restrictions used in the analysis, wherein charged pions less than 150 MeV/$c$ momentum may be included within the final states of absorption and charge exchange, and multiple $\pi^0$s can be present in charge exchange. Fig.~\ref{fig:corr_xsec} shows the correlations between cross section bins defined as $c_{ij} = \frac{v_{ij}}{\delta_i\delta_j}$. Here, $v_{ij}$ is the covariance between bins $i$ and $j$, and $\delta_{i(j)}$ is the square root of the variance of bin $i$ ($j$) and is shown as the error bars in Figs.~\ref{fig:abs_xsec} -- \ref{fig:other_xsec}.
Tab.~\ref{tab:xsecs} summarizes the measured cross sections and their variances. The $\chi^2$ shown in~\ref{fig:abs_xsec} is defined as $\chi^2 = (\vec{\sigma} - \vec{\sigma}_{\textrm{pred}})^TV^{-1}(\vec{\sigma} - \vec{\sigma}_{\textrm{pred}})$, where $\vec{\sigma}$ are the measured cross sections, $\vec{\sigma}_{\textrm{pred}}$ are the predicted cross sections in question, and $V^{-1}$ is the inverted covariance matrix. 

Of the models shown here, the data is consistent with the Bertini cascade and INCL++ models from \geant~as well as GENIE's \hN~model, with a slight preference for the Bertini cascade ($p\approx0.31$, $0.08$, $0.10$ respectively)\footnote{9 degrees of freedom are used within the $\chi^2$ calculations.}. Meanwhile, GENIE's \hA~model is inconsistent with this data ($p <$ 0.001). This is not surprising, as \hA~is an effective model, rather than a full intranuclear cascade model like the other models.

\begin{figure*}[!htb]
  \subfloat[Total Inelastic Cross Section\label{fig:total_xsec}]{
    \includegraphics[width=.32\textwidth]{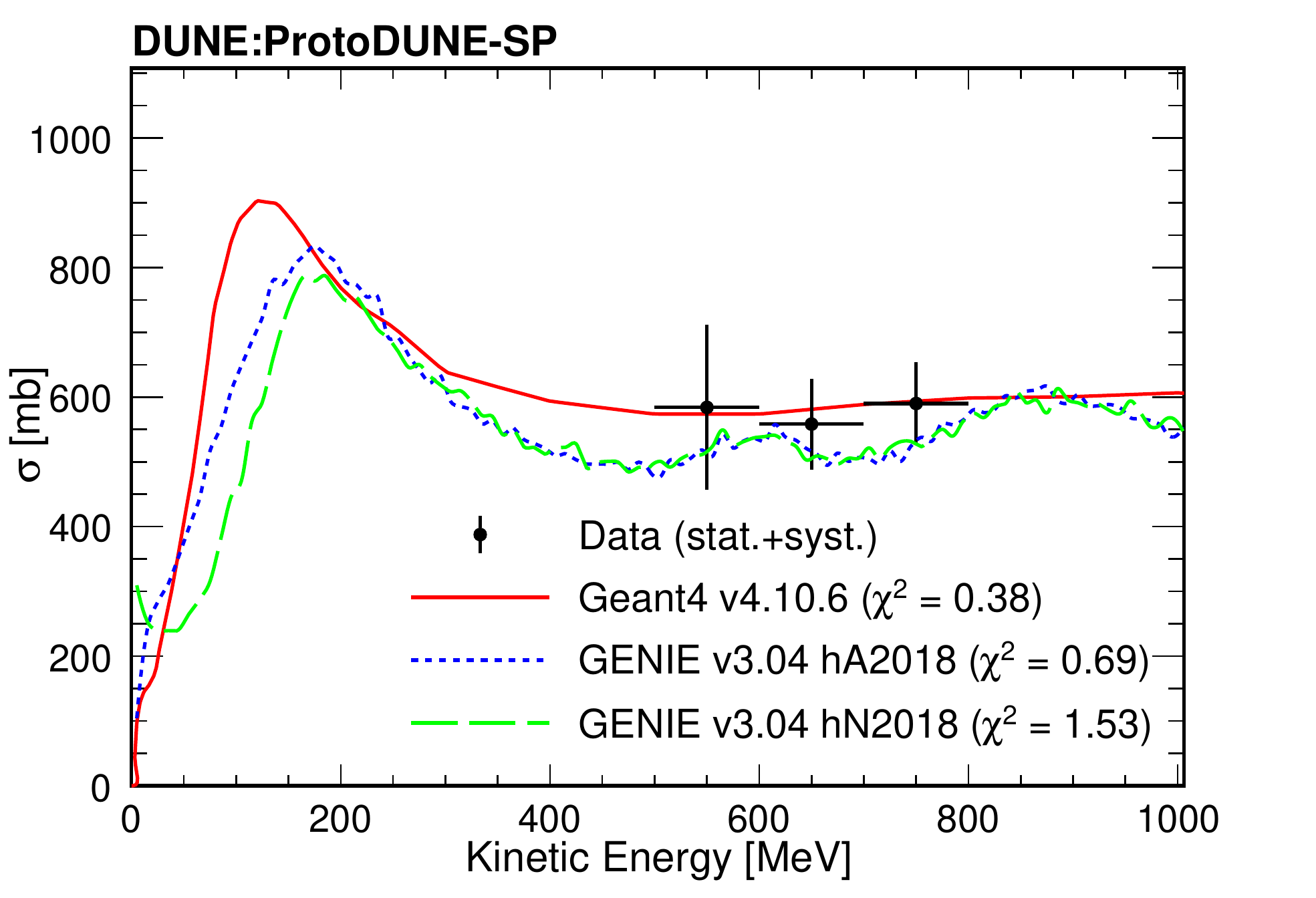}
  }
  \subfloat[Correlation Matrix\label{fig:corr_total_xsec}]{
    \includegraphics[width=.32\textwidth]{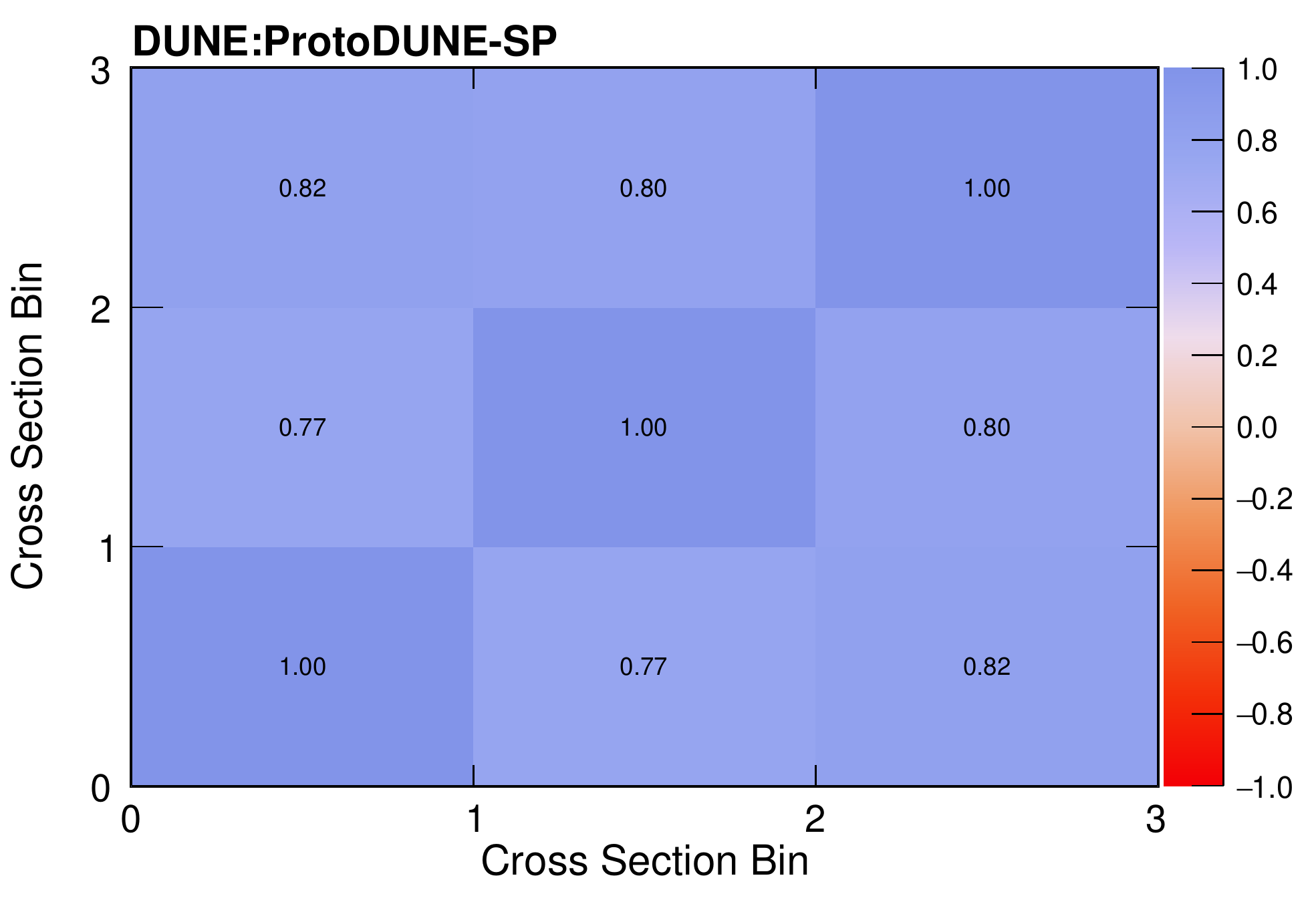}
  }
  \caption{~\ref{fig:total_xsec} ProtoDUNE-SP measurement of the total cross section compared to \geant~and \genie~predictions. \ref{fig:corr_total_xsec} shows the correlations between the data points shown here in order of increasing kinetic energy.}
  \label{fig:results_xsecs_total}
\end{figure*}

We also present the total inelastic cross section in Fig.~\ref{fig:results_xsecs_total} calculated by summing the measured exclusive cross sections. It is important to stress that this measurement of the total cross section is not independent from the measurement of the three exclusive channels, but is statistically valid for standalone comparisons to predicted total inelastic \piplus cross sections. The covariance matrix of the total cross section is calculated as:
\begin{equation}
    V_{ij} = \sum\limits_{\mu}^{\{a, b, c\}}\sum\limits_{\nu}^{\{a, b, c\}}v_{\mu_i \nu_j}
\end{equation} 
where $V_{ij}$ is the covariance of bins $i$ and $j$ in the total cross section. $v_{\mu_i \nu_j}$ is the covariance between bin $i$ of exclusive channel $\mu$ and bin $j$ of exclusive channel $\nu$. The correlation matrix shown in Fig.~\ref{fig:corr_total_xsec} is defined as $C_{ij} = \frac{V_{ij}}{\Delta_i\Delta_j}$ where the variance, $\Delta_i$, is the $i^{\textrm{th}}$ element of the diagonal of $V$. As can be seen, the bins of the measured total inelastic cross section are highly correlated.

We compare the total inelastic cross section to predictions from \geant~ and GENIE's $hA$ and $hN$ models in Fig.~\ref{fig:total_xsec}. Note that only one \geant~ prediction is shown, as the total cross section is out of scope of the Bertini and INCL cascade models (they only provide the dynamics of inelastic pion scattering interactions, not the rate at which the scattering occurs~\cite{geant4phys}). Again, the \geant~ prediction agrees best with the data, though each model is consistent with the data as indicated by the $\chi^2$ values shown in the figure. A dedicated analysis on the total inelastic cross section is currently underway and a publication is being prepared. The results from the total inelastic analysis and the combined cross sections presented here are consistent.

\section{Conclusions}
We have measured the absorption, charge exchange, and remaining inelastic $\pi^+$--argon cross sections using data taken by the ProtoDUNE-SP detector. This analysis showcases the capability of the DUNE LArTPC detector technology to perform particle tracking, identification, and energy reconstruction. This measurement will provide necessary inputs for the simulation used by DUNE and will improve the modeling of pions undergoing final state interactions within the nucleus as well as secondary interactions with the LAr volume of the DUNE detectors following a neutrino interaction. It is the first ever measurement of $\pi^+$--argon charge exchange and the first measurement of absorption on argon at the kinetic energy range of 500--800 MeV.


\begin{acknowledgments}

%
%
The ProtoDUNE-SP detector was constructed and operated on the CERN Neutrino Platform.
We gratefully acknowledge the support of the CERN management, and the
CERN EP, BE, TE, EN and IT Departments for NP04/Proto\-DUNE-SP.
%
%
This document was prepared by DUNE collaboration using the resources of the Fermi National Accelerator Laboratory (Fermilab), a U.S. Department of Energy, Office of Science, Office of High Energy Physics HEP User Facility. Fermilab is managed by Fermi Forward Discovery Group, LLC, acting under Contract No. 89243024CSC000002.
%
%
This work was supported by
CNPq,
FAPERJ,
FAPEG and 
FAPESP,                         Brazil;
CFI, 
IPP and 
NSERC,                          Canada;
CERN;
M\v{S}MT,                       Czech Republic;
ERDF, 
Horizon Europe, 
MSCA and NextGenerationEU,      European Union;
CNRS/IN2P3 and
CEA,                            France;
INFN,                           Italy;
FCT,                            Portugal;
NRF,                            South Korea;
Generalitat Valenciana, 
Junta de AndalucÄ±a-FEDER, 
MICINN, and 
Xunta de Galicia,               Spain;
SERI and 
SNSF,                           Switzerland;
T\"UB\.ITAK,                    Turkey;
The Royal Society and 
UKRI/STFC,                      United Kingdom;
DOE and 
NSF,                            United States of America.
%
%
This research used resources of the 
National Energy Research Scientific Computing Center (NERSC), 
a U.S. Department of Energy Office of Science User Facility 
operated under Contract No. DE-AC02-05CH11231.
\end{acknowledgments}


\appendix

\bibliography{thebib}
\end{document}